\documentclass[acmtog,nonacm,authorversion]{acmart}
\acmSubmissionID{251}
\usepackage{subcaption}
\usepackage{algorithm}
\usepackage{algpseudocode}
\usepackage{graphicx}
\usepackage{wrapfig}
\usepackage{subcaption}

\AtBeginDocument{%
  \providecommand\BibTeX{{%
    \normalfont B\kern-0.5em{\scshape i\kern-0.25em b}\kern-0.8em\TeX}}}
    
\makeatletter
\renewcommand*\env@matrix[1][\arraystretch]{%
  \edef\arraystretch{#1}%
  \hskip -\arraycolsep
  \let\@ifnextchar\new@ifnextchar
  \array{*\c@MaxMatrixCols c}}
\makeatother

\begin{document}

\title{A Neural-Network-Based Approach for Loose-Fitting Clothing}

\author{Yongxu Jin}
\email{yxjin@stanford.edu}
\affiliation{%
  \institution{Stanford University}
  \city{Stanford}
  \state{California}
  \country{USA}
}
\affiliation{%
  \institution{Epic Games}
  \city{Cary}
  \state{North Carolina}
  \country{USA}
}

\author{Dalton Omens}
\email{domens@stanford.edu}
\affiliation{%
  \institution{Stanford University}
  \city{Stanford}
  \state{California}
  \country{USA}
}
\affiliation{%
  \institution{Epic Games}
  \city{Cary}
  \state{North Carolina}
  \country{USA}
}

\author{Zhenglin Geng}
\email{zhenglin.geng@epicgames.com}
\affiliation{%
  \institution{Epic Games}
  \city{Cary}
  \state{North Carolina}
  \country{USA}
}

\author{Joseph Teran}
\email{jteran@math.ucdavis.edu}
\affiliation{%
  \institution{University of California, Davis}
  \city{Davis}
  \state{California}
  \country{USA}}
\affiliation{%
  \institution{Epic Games}
  \city{Cary}
  \state{North Carolina}
  \country{USA}
}

\author{Abishek Kumar}
\email{Abishek.Kumar@sony.com}
\affiliation{%
  \institution{Sony Corporation of America, R\&D US Laboratory}
  \country{USA}
}

\author{Kenji Tashiro}
\email{Kenji.Tashiro@sony.com}
\affiliation{%
  \institution{Sony Corporation of America, R\&D US Laboratory}
  \country{USA}
}

\author{Ronald Fedkiw}
\email{rfedkiw@stanford.edu}
\affiliation{%
  \institution{Stanford University}
  \city{Stanford}
  \state{California}
  \country{USA}
}
\affiliation{%
  \institution{Epic Games}
  \city{Cary}
  \state{North Carolina}
  \country{USA}
}


\begin{abstract}
Since loose-fitting clothing contains dynamic modes that have proven to be difficult to predict via neural networks, we first illustrate how to coarsely approximate these modes with a real-time numerical algorithm specifically designed to mimic the most important ballistic features of a classical numerical simulation. 
Although there is some flexibility in the choice of the numerical algorithm used as a proxy for full simulation, it is essential that the stability and accuracy be independent from any time step restriction or similar requirements in order to facilitate real-time performance.
In order to reduce the number of degrees of freedom that require approximations to their dynamics, we simulate rigid frames and use skinning to reconstruct a rough approximation to a desirable mesh; 
as one might expect, neural-network-based skinning seems to perform better than linear blend skinning in this scenario. 
Improved high frequency deformations are subsequently added to the skinned mesh via a quasistatic neural network (QNN).
In contrast to recurrent neural networks that require a plethora of training data in order to adequately generalize to new examples, QNNs perform well with significantly less training data.  

\end{abstract}



\begin{teaserfigure}
    \includegraphics[width=\textwidth]{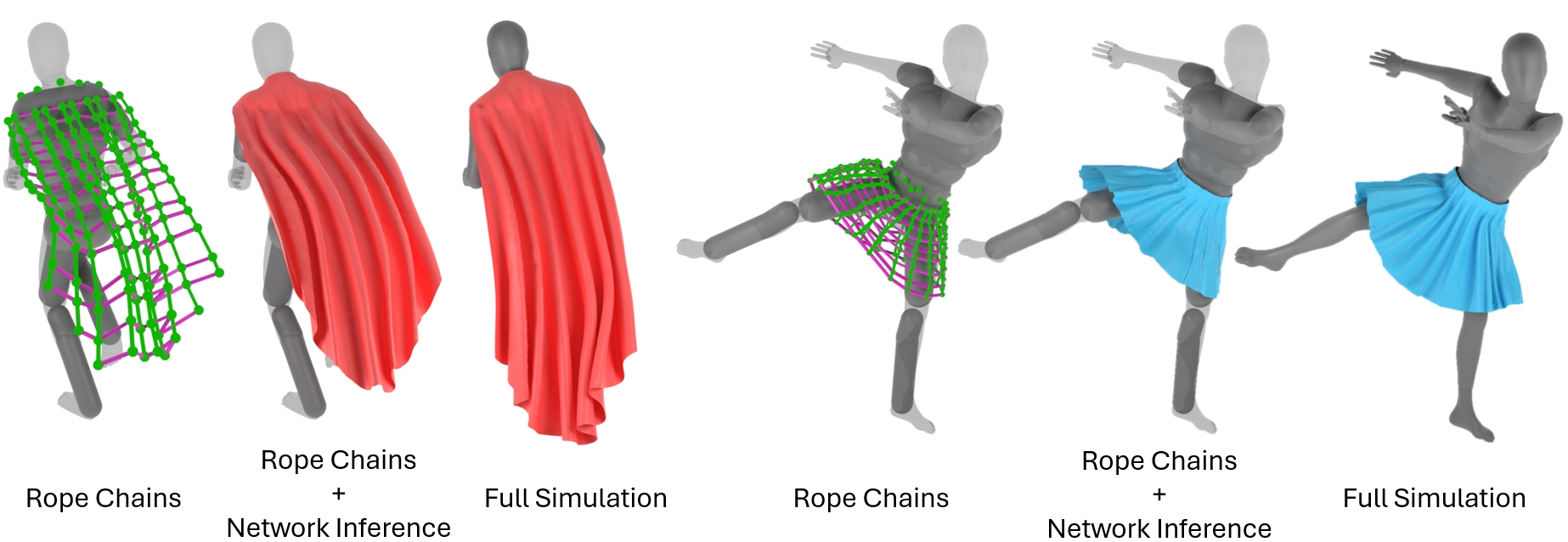}
    \caption{We introduce a rope chain simulation approach to efficiently model cloth dynamics using a small number of degrees of freedom. Analytic signed distance functions are used to efficiently manage collisions with the body mesh. Neural networks are utilized to skin a mesh from the simulated degrees of freedom and to capture the detailed mesh shapes. Our results (the second and fifth images) not only produce dynamics similar to full numerical simulations (the third and sixth images) but also do not suffer from the locking and/or overstretching typical of real-time physics-based simulations.}
\end{teaserfigure}

\maketitle

\section{Introduction}
\label{sec:intro}

Animation of digital clothing has been a captivating research area for decades due to its importance in creating realistic digital humans. 
While traditional physics-based simulation methods \cite{baraff1998large, bridson2002robust, choi2002stable} can generate high-fidelity results, the inability to achieve real-time performance at high resolutions restricts their utility in contemporary real-time applications such as video games, VR/AR, and virtual try-on systems. 
In light of recent advancements in GPU hardware, there has been a surge of interest in leveraging neural networks to approximate physics-based simulations, see e.g. \cite{raissi2019physics, santesteban2019learning, patel2020tailornet, sanchez2020learning, lewin2021swish,  pfaff2021learning}

Tight or close fitting garments such as shirts, pants, etc. typically exhibit only subtle dynamic behaviors; thus, capturing quasistatic shape information is more important than modelling the ballistic motion, see e.g. \cite{lahner2018deepwrinkles, jin2020pixel, patel2020tailornet,bertiche2021pbns}. 
Quasistatic shape information can be captured by various skinning techniques, by neural networks that lack temporal information (we will refer to these as quasistatic neural networks or QNNs), or by a combination of skinning and quasistatic neural network approaches.
While some low-frequency vibration of tight-fitting clothing can be captured by underlying flesh motion (see e.g. \cite{jin2022analytically}), modelling the pronounced ballistic motion associated with loose-fitting clothing such as skirts, dresses, capes, etc. is more challenging.
Skinning and QNN based approaches have not been able to successfully model pronounced ballistic motion; thus, researchers have explored alternative network architectures that incorporate temporal history, most notably recurrent neural networks (RNNs), see e.g. \cite{santesteban2022snug, zhang2021dynamic, pan2022predicting}. Unfortunately, the amount of training data required to robustly model temporal transitions between states (especially when considering the need for generalization) is significantly greater than that required to model the states themselves.
Moreover, increased network capacity is required in order to properly capture the numerous state transitions present in this increased volume of training data.
These issues typically cause recurrent neural network approaches to overfit and thus generalize poorly.

Since a physics-based simulation readily models ballistic motions but does not efficiently scale to a large number of degrees of freedom and neural networks readily deal with a large number of degrees of freedom but struggle with temporal state transitions, we pursue a more optimal hybrid approach that uses a small number of degrees of freedom physics simulation to capture dynamics and a neural network to capture a high resolution shape. In particular, we physically simulate such a low number of degrees of freedom that they are best viewed as virtual bones as opposed to being viewed as a subset of a higher resolution mesh (similar to \cite{pan2022predicting, zhao2023learning}); thus, we rely on skinning (driven by a neural network) in order to construct a coarse approximation to the desired mesh from the simulated degrees of freedom. Given this coarsely approximated dynamic mesh, a quasistatic neural network is then used to obtain a more desirable higher resolution mesh.

Since we rely on the neural network's ability to capture the shape of the cloth, the physics simulation does not require all the usual (and computationally expensive) techniques for simulating stretching, bending, compression, etc.; thus, we devise a novel approach that connects our virtual bones (or rigid frames) into vertical rope chains.
Each rope is designed to be inextensible yet shrinkable, avoiding undesirable locking or rubbery buckling artifacts (see \cite{jin2017inequality} for detailed insights). 
Optional spring forces can be included in order to  softly constrain the distances between different rope chains and/or to regulate the shrinking of each rope (if desired).
Compared to low resolution cloth simulation, the rope chains can be robustly simulated with large time steps thus enabling real-time performance.
Instead of simulating the rotational (in additional to translational) degrees of freedom typically required in order to skin a coarse mesh from virtual bones (or rigid frames), we utilize a neural network so that the garment can be skinned directly from the simulated translational degrees of freedom. 

Collisions between the body and the rope chain degrees of freedom are facilitated via signed distance functions (SDFs), see e.g. \cite{bridson2003simulation}. 
In order to avoid computationally expensive grid-based SDFs, the SDF can be defined either by a set of closed-form primitives (desirable for real-time applications like video games) or by a neural network (see e.g. \cite{park2019deepsdf, remelli2020meshsdf}).
Similar collision treatment can also be applied to the degrees of freedom of the full cloth mesh. Both the skinning neural network and the QNN are trained with an additional PINN-style \cite{raissi2019physics} collision loss (using the SDF) in order to obtain network parameters that favor collision-free cloth mesh degrees of freedom; importantly, adding collisions in this fashion does not require modifications to the network architecture nor does it add any computational expense to inference. 

To summarize our contributions:
\begin{itemize}
    \item We propose a hybrid framework for animating loose-fitting clothing that blends together the efficacy of physics simulation for capturing ballistic motion and the efficiency of neural networks for skinning and shape inference. 
    \item In particular, we propose a novel simulation method for low resolution (and loose-fitting) clothing via ballistic rope chains, which are used to reconsturct a full cloth mesh with the aid of neural networks for both skinning and shape inference.
    \item We propose a novel collision handling method with analytic SDFs, using history-based information in order to improve both robustness and efficiency for the sake of real-time applications.
\end{itemize}

\section{Related Work}

\paragraph{\textbf{Physics Simulation:}}
The physical simulation of cloth has a long hisory in computer graphics, dating back to \cite{terzopoulos1987elastically}; however, it gained significant popularity due to the implicit time integration approach proposed in \cite{baraff1998large}. In order to overcome the overly-damped (underwater) appearance of implicit time integration, \cite{bridson2002robust, bridson2003simulation} introduced a semi-implicit time integration approach (central differencing in computational mechanics) that is explicit on the elastic vibrational modes and implicit on the damping modes. Although offline methods could simulate very high resolution cloth even 15 years ago (see \cite{selle2008robust}), position based dynamics (starting with \cite{muller2007position}) has been the method of choice for most real-time applications.
Other interesting contributions include discussions on the buckling instability \cite{choi2002stable}, overcoming locking \cite{jin2017inequality}, etc.

\paragraph{\textbf{Neural Physics:}}
Many researchers have aimed to mimic physics simulations via neural networks and various other data-driven techniques. 
Early works include: \cite{de2010stable, hahn2014subspace} used a PCA subspace, \cite{guan2012drape} predicted a cloth mesh from pose history and body shape, \cite{kim2013near} used motion graphs, and \cite{pons2015dyna,loper2015smpl} used linear auto-regression.
As deep learning gained popularity, a number of authors embraced neural networks.
\cite{luo2018nnwarp} used a neural network to add non-linear displacements on top of linear elasticity.
\cite{holden2019subspace} used a neural network in a PCA subspace to replace time integration.
\cite{fulton2019latent, tan2020realtime, wang2019learning} all aimed to integrate state transitions in a latent space (see also \cite{zhang2021dynamic}).
\cite{sanchez2020learning, chentanez2020cloth, pfaff2021learning, d2022n} all used graph neural networks, which can embrace the typical physics based simulation notion of a stencil (see also \cite{zheng2021deep}).
\cite{santesteban2022snug, grigorev2023hood} used a PINN-style objective function.
Although PINNs (first proposed in \cite{raissi2019physics}) are self-supervised and as such do not require ground truth data, there is nothing special about their architecture implying that they need just as much training data as any other method in order to properly generalize to unseen data.
\cite{shao2023towards} used Transformers. 
\cite{li2023d} used three networks to capture static, coarse, and wrinkle dynamics separately.
Most similar to our approach, \cite{pan2022predicting, zhao2023learning, diao2023combating} used virtual bones, skinning, and/or super-resolution techniques; however, (unlike us) they used RNNs to animate the virtual bones.

\paragraph{\textbf{Rigging and Skinning:}}
There is a long history of rigging and skinning in computer graphics, although not necessarily geared towards cloth animation. 
We refer interested readers to \cite{magnenat1988joint} for linear blend skinning, \cite{lewis2000pose} for pose space deformation, \cite{kurihara2004modeling} for weighted pose space deformation, \cite{kavan2007skinning, kavan2008geometric} for dual quaternion skinning, and \cite{rumman2017skin} for a survey.
Other interesting works include the parametric body models in \cite{anguelov2005scape} (SCAPE) and \cite{loper2015smpl} (SMPL) and the joint extraction in \cite{le2012smooth} (SSDR).
It is difficult to skin clothing (especially when it is loose-fitting) using standard (non-neural) skinning techniques; however, see e.g. \cite{wang2010example, xu2014sensitivity}.

\paragraph{\textbf{Neural Shape:}}
There have been a number of efforts to infer shape (and appearance, see \cite{lahner2018deepwrinkles}) using neural networks.
\cite{bailey2018fast} added per-vertex displacements on top of the skinned body mesh, and \cite{bailey2020fast} similarly added per-vertex displacements on top of a skinned face mesh.
\cite{jin2022analytically} also added per-vertex displacements on top of skinned body mesh, but augmented these with dynamic motion from analytic springs.
For cloth, there have been various attempts to create a high resolution mesh from the physics simulation of a coarser mesh, see e.g. \cite{kavan2011physics, oh2018hierarchical, chentanez2020cloth}.
Inferring cloth from body pose and shape alone (without a coarse simulation mesh) is significantly more difficult, see e.g. \cite{santesteban2019learning, gundogdu2019garnet, patel2020tailornet, jin2020pixel, lewin2021swish, bertiche2021pbns, tan2022repulsive, lee2023clothcombo, li2024isp}.

\paragraph{\textbf{Collisions:}}
The successful approach to cloth-cloth self collisions using continuous collision detection (CCD) in \cite{bridson2002robust} (which leveraged \cite{provot1997collision}) led to a plethora of work on improving the efficiency of $O(n\log n)$ collision detection (see e.g. \cite{govindaraju2005interactive, sud2006fast, tang2014fast, tang2018cloth, li2020p}); however, for cloth-body collisions, signed distance functions (SDFs) remain prevalent due to their more efficient $O(1)$ cost (see e.g. \cite{bridson2003simulation}).
The main drawback of SDFs is that the three dimensional discretization of the volumetric field is expensive to load and store in memory; thus, analytic SDFs are more popular for real-time applications (see e.g. \cite{blinn1982generalization, wyvill1986soft, nishimura1985object}). 
Neural network approximations to SDFs have the potential to be efficient enough to be used in real-time applications. Inference is typically fast enough, but the amount of network parameters required (or new network parameters required, when switching from one SDF to another) needs to be small enough to be efficiently loaded and/or stored in memory. We refer the interested reader to \cite{saito2019pifu, park2019deepsdf, remelli2020meshsdf, shen2021deep, mehta2022level, wu2023weakly} for various details (and discussions on differentiability).

\paragraph{\textbf{Physics-Informed Neural Networks (PINNs):}}
For neural network inferenced cloth, the computational burden from processing collisions can be alleviated to some degree by utilizing interpenetration-free training data. 
Unfortunately, regularization (which is generally necessary and desirable) prevents inferenced cloth from being interpenetration-free even when the training data is interpenetration-free.
Thus, PINN-style \cite{raissi2019physics} collision losses can be quite useful during training (see \cite{santesteban2021self, bertiche2021pbns, bertiche2021deepsd, gundogdu2019garnet, santesteban2022snug}), as they allow one to increase the penalty on cloth-body interpenetrations without forcing overfitting to the interpenetration-free training data. The PINN losses can be made as stiff as desired while otherwise maintaining desirable regularization on the deviation of the cloth from the training data positions.

\begin{figure}
    \includegraphics[width=\columnwidth]{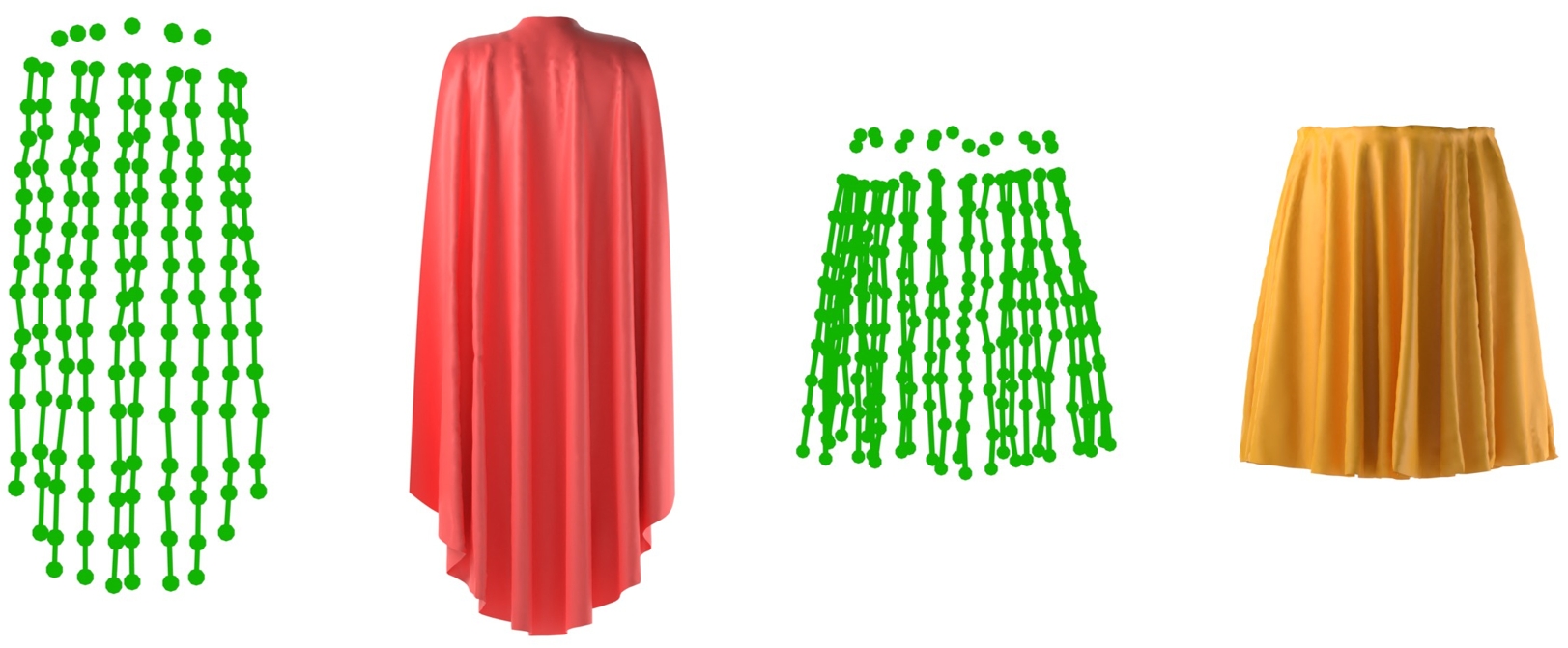}
  \caption{The loose-fitting garments that we use for numerical experiments. The cape mesh consists of 12690 vertices, and we define 10 rope chains with 13-15 virtual bones in each chain (reducing the total DOFs by a factor of approximately 90). The skirt mesh consists of 18546 vertices, and we define 26 rope chains with 9 virtual bones in each chain (reducing the total DOFs by a factor of approximately 80). Importantly, the large reduction in the number of DOFs is highly beneficial for both RAM and cache performance, not just CPU performance. Note that the unattached virtual bones near the top of both the cape and the skirt are not simulated, but they will be used for skinning (see Section \ref{sec:skinning}).}
  \label{fig:pendsetup}
\end{figure}

\section{Rope Chain Simulation}
\label{sec:ropechainsim}

\begin{figure}
    \includegraphics[width=\columnwidth]{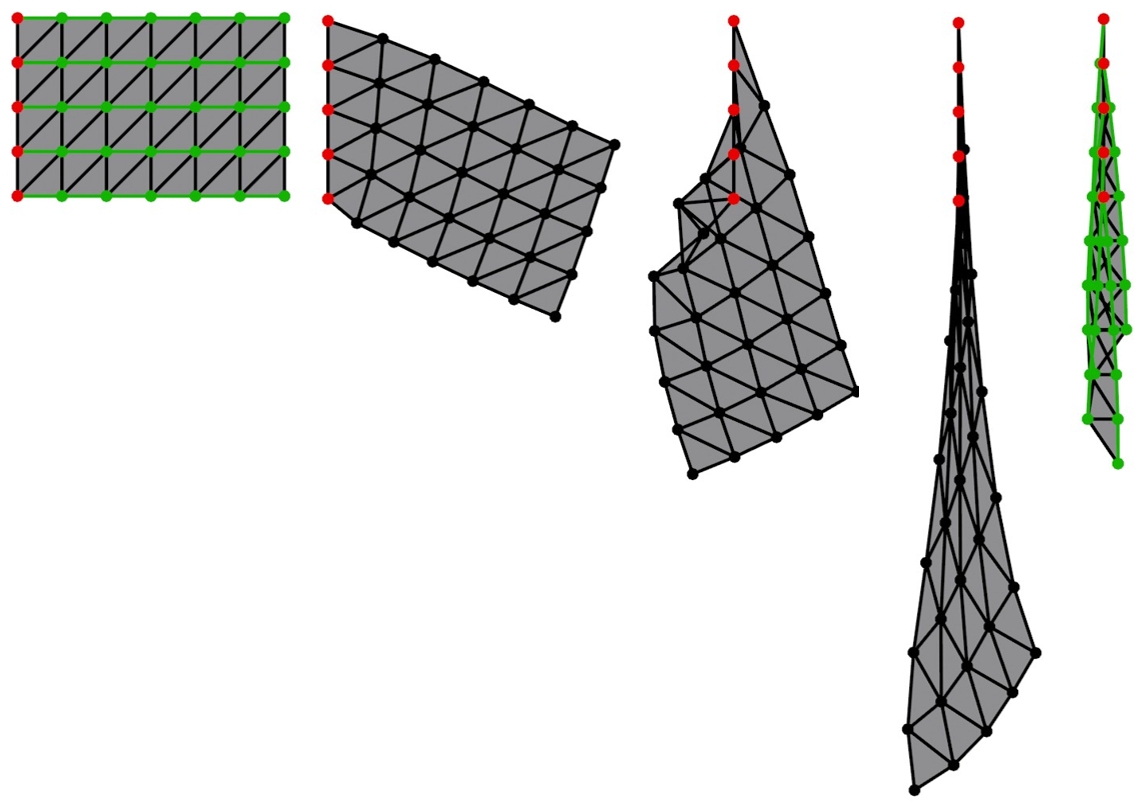}
  \caption{The far left subfigure shows a a coarsely discretized mesh with position constraints on the five red nodes. The next three subfigures show the results of steady state simulations using a decreasing spring stiffness (from left to right). The final subfigure shows a rope chain simulation (the rope chains are shaded green) of the same degrees of freedom. The mass-spring simulations lock with stiffer springs and overstrech with weaker springs. The spring stiffness in the middle subfigure was chosen to approximately match the downward stretching extent of the rope chain simulation, which is what one would expect without overstretching; however, locking occurs since the springs are still too stiff.}
  \label{fig:flag}
\end{figure}

Given a garment mesh, we define a set of virtual bones distributed across the garment (either manually or by a procedural algorithm such as SSDR \cite{le2012smooth});
then, the virtual bones are interconnected vertically to form a set of simulatable rope chains (see Figure \ref{fig:pendsetup}). 
Not only does this significantly reduce the number of degrees of freedom that need to be simulated, but (we argue that) the rope chains provide a much better approximation to the desired ballistic degrees of freedom than a rubbery mass-spring mesh does: Rope chains can be made to swing and rotate freely, whereas a mass-spring system follows linearized rotations resisted by spring stretching. 
Rope chains can be made to buckle freely, whereas a mass-spring system over-resists buckling (locking when under-discretized, see e.g. \cite{jin2017inequality}). 
Enforcing inextensibility is trivial (from root to tip) for a rope chain, whereas various ad-hoc techniques are required in order to prevent over-stretching for a mass-spring simulation. Etc. See Figure \ref{fig:flag}.

\begin{figure}
    \includegraphics[width=\columnwidth]{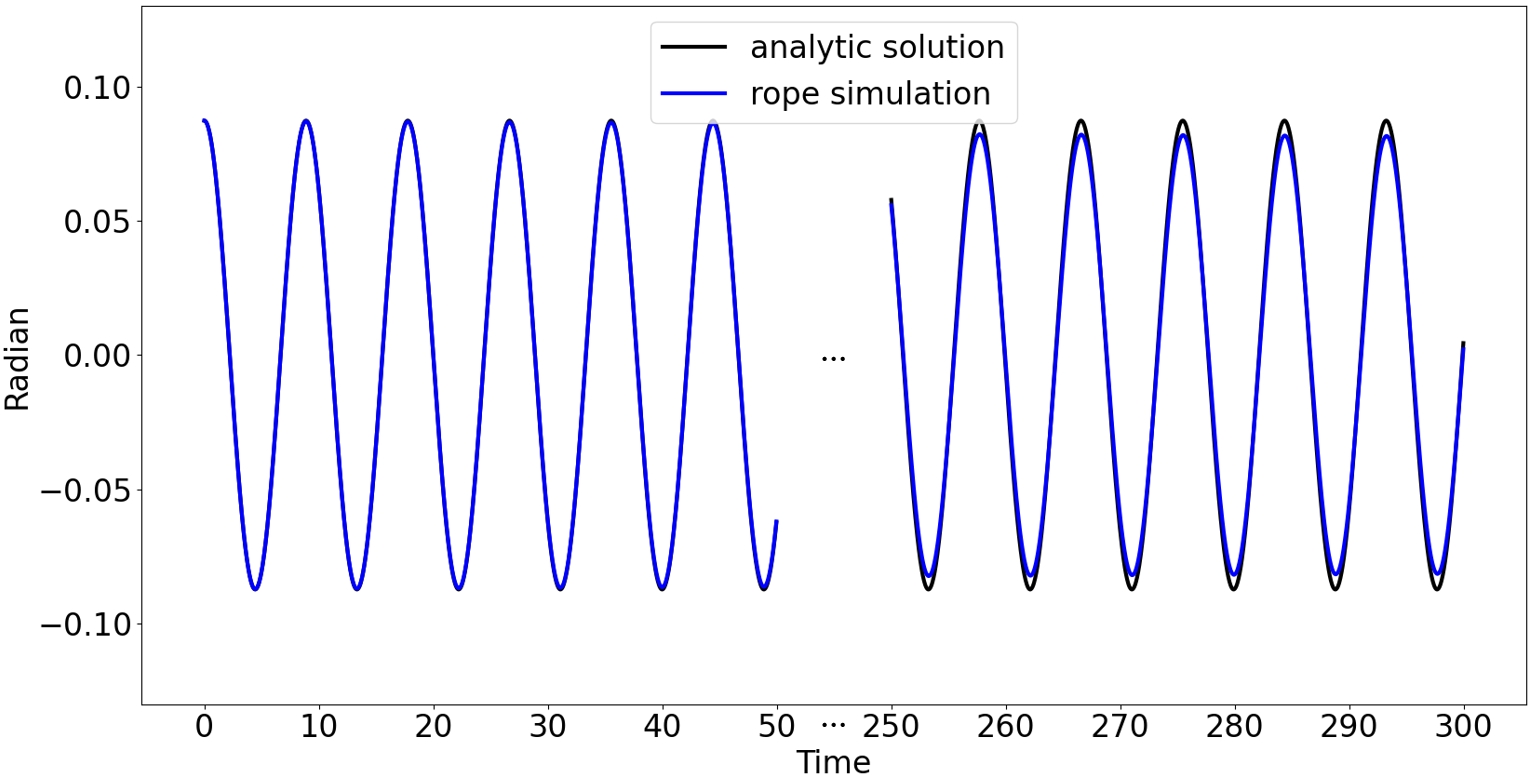}
    \caption{In this figure, we simulate two virtual bones connected by a single rope with the top virtual bone fixed and the bottom virtual bone free to rock back and forth as a pendulum. The results compare well to the analytic solution for pendulum motion for both shorter times (left) and longer times (right), illustrating the efficacy of our numerical approach.}
    \label{fig:analy}
\end{figure}

A semi-implicit Newmark style time integration scheme is used separately for each rope chain (see e.g. \cite{bridson2003simulation}):
\begin{enumerate}
    \item $\vec{v}^{n + \frac{1}{2}} =  \vec{v}^{n} + \frac{\Delta t}{2}  \frac{\vec{F} (\vec{x}^n, \vec{v}^n)}{M}$ 
    \item $\vec{x}^{n + 1} =\vec{x}^{n} + \Delta t  \vec{v}^{n + \frac{1}{2}}$
    \item Resolve collisions, perturbing $\vec{x}^{n + 1}$ and $\vec{v}^{n + \frac{1}{2}}$
    \item $\vec{v}^{n + 1} =  \vec{v}^{n + \frac{1}{2}} + \frac{\Delta t}{2}  \frac{\vec{F} (\vec{x}^{n+1}, \vec{v}^{n+\frac{1}{2}})}{M}$ 
\end{enumerate}
This is oftern referred to as central differencing in the computational mechanics literature. Central differencing preserves rich high-frequency dynamic motion significantly better than fully-implicit methods (such as backward Euler) do. The velocity updates (steps 1 and 4) are discussed in detail in Section \ref{sec:velupdate}, the position update (step 2) is discussed in detail in Section \ref{sec:posupdate}, and collisions (step 3) are discussed in detail in Section \ref{sec:collision}. See Figure \ref{fig:analy}.

\section{Velocity Update}
\label{sec:velupdate}

Given a rope chain, let ${\vec{x}_0, ..., \vec{x}_m}$ denote the virtual bones from root to tip where $\vec{x}_0$ is kinematically constrained to follow some part of the body (or some other object). 
The magnitude of each $\vec{l}_i=\vec{x}_i - \vec{x}_{i-1}$ should never exceed the maximal length $l_{max, i}$ of the corresponding rope; however, there is no penalty for slack ($|\vec{l}_i|=l_i<l_{max, i}$) in the rope. Except for $\vec{x}_0$, various external forces $\vec{F}_{ext, i}$ are applied to the virtual bones. Examples include gravity $\vec{F}_{g} = M_i \vec{g}$, wind drag $\vec{F}_{wind} = -c_{wind} (\vec{v}_i - \vec{v}_{wind})$, etc. In order to prevent a rope chain from deviating too far from its neighboring rope chains, springs can be attached laterally connecting each virtual bone to its neighbors on neighboring rope chains. We treat these as soft constraints, meant to influence but not overly dictate the simulation; thus, they are added to $\vec{F}_{ext, i}$ and given the same relative importance as gravity, wind drag, and other similar forces. One may also desire forces that aim to preserve rest angles between consecutive pairs of virtual bones, in order to coerce the cloth towards its rest shape. These forces would also be included in $\vec{F}_{ext, i}$.

When a rope reaches its maximal length $l_{max, i}$, the two virtual bones it attaches to will rotate around each other.
The magnitude of the centripetal force required to maintain this rotation is 
\begin{align}
    \label{eqn:centripetal}
    F_{c, i} &= M_{i-1} \frac{|(\vec{v}_{i-1}  - \vec{v}_c) -\left((\vec{v}_{i-1}  - \vec{v}_c) \cdot \hat{l}_i\right) \hat{l}_i |^2}{|\vec{x}_{i-1} - \vec{x}_c|} \nonumber \\
    &= M_i \frac{|(\vec{v}_i - \vec{v}_c) - \left((\vec{v}_i - \vec{v}_c) \cdot \hat{l}_i\right) \hat{l}_i|^2}{|\vec{x}_i - \vec{x}_c|}
\end{align}
where $\vec{x}_c=\frac{M_{i-1}\vec{x}_{i-1} + M_{i}\vec{x}_{i}}{M_{i-1} + M_i}$ is the center of mass, $\vec{v}_c=\frac{M_{i-1}\vec{v}_{i-1} + M_{i}\vec{v}_{i}}{M_{i-1} + M_i}$ is the velocity at the center of mass, and $\hat{l}_i = \frac{\vec{l}_i}{l_i}$ is unit length (i.e. a direction).
For the first rope, which connects the kinematic $\vec{x}_0$ with $\vec{x}_1$, $\vec{x}_c=\vec{x}_0$ and $\vec{v}_c=\vec{v}_0$ due to $M_0=\infty$ (note that second line of Equation \ref{eqn:centripetal} needs to be used to calculate $F_{c, 1}$, in order to avoid dealing with L'Hopital's rule in the first line).

When a rope reaches its maximal length $l_{max, i}$, an additional tension force with magnitude $T_i \ge 0$ is added to the non-kinematic virtual bones it attaches to (the kinematic root ignores these forces). The net force in the virtual bones can be defined via
\begin{subequations}
\label{eqn:netforce_all}
\begin{align}
    \label{eqn:netforce}
    \vec{F}_{net, i} &= \vec{F}_{ext, i} - T_{i} \hat{l}_{i} + T_{i+1} \hat{l}_{i+1} \\
    \label{eqn:netforce_m}
    \vec{F}_{net, m} &= \vec{F}_{ext, m} - T_{m} \hat{l}_{m}
\end{align}
\end{subequations}
where $i \in \{1, ..., m-1\}$ and $T_i=0$ whenever $l_i < l_{max, i}$. 
In order to preserve the rotational motion for each taut rope,
\begin{subequations}
\begin{align}
    \label{eqn:netineq_1}
    \vec{F}_{net, 1} \cdot \hat{l}_1 &\le -F_{c, 1} + M_1 \ddot{\vec{x}}_0 \cdot \hat{l}_1 \\
    \label{eqn:netineq}
    \vec{F}_{net, i} \cdot \hat{l}_i - \vec{F}_{net, i-1} \cdot \hat{l}_i &\le -2F_{c, i} 
\end{align}
\end{subequations}
where $i \in \{2, ..., m\}$ and the $\ddot{\vec{x}}_0$ term accounts for the motion of the kinematic root.
Substituting Equations \ref{eqn:netforce} and \ref{eqn:netforce_m} into Equations \ref{eqn:netineq_1} and \ref{eqn:netineq} gives
\begin{subequations}
\label{eqn:tension_set}
\begin{align}
    - T_1 + \hat{l}_2 \cdot \hat{l}_1 T_2 \le -  \vec{F}_{ext, 1} \cdot \hat{l}_1 -F_{c, 1} + M_1 \ddot{\vec{x}}_0 \cdot \hat{l}_1 \label{eqn:tension_1} \\
    \hat{l}_{i-1} \cdot \hat{l}_i T_{i-1} - 2T_i +  \hat{l}_{i+1} \cdot \hat{l}_i T_{i+1} \le (\vec{F}_{ext, i-1}- \vec{F}_{ext, i}) \cdot \hat{l}_i - 2F_{c, i} \label{eqn:tension} \\
    \hat{l}_{m-1} \cdot \hat{l}_m T_{m-1} - 2T_m  \le (\vec{F}_{ext, m-1}- \vec{F}_{ext, m}) \cdot \hat{l}_m - 2F_{c, m} \label{eqn:tension_m}
\end{align}
\end{subequations}
where $i \in \{2, ..., m-1\}$. This tri-diagonal linear system of inequalities (for the unknown tension) decouples into separate blocks whenever a slack rope has $l_i < l_{max, i}$ or two consecutive ropes are orthogonal (with a dot product of zero). For computational efficiency, we iterate these equations in tip-to-root order (from bottom to top in Equation \ref{eqn:tension_set}) using Gauss-Seidel; for most real-time applications, typically only one tip-to-root sweep is required. In particular, we rewrite Equations \ref{eqn:tension_1}, \ref{eqn:tension}, and \ref{eqn:tension_m} in reverse order as
\begin{subequations}
\label{eqn:gaussseidel_set}
\begin{align}
    T_m  &\ge \frac{(\vec{F}_{ext, m} - \vec{F}_{ext, m-1}) \cdot \hat{l}_m + 2F_{c, m} + \hat{l}_{m-1} \cdot \hat{l}_m T_{m-1}}{2} \\
     T_i  &\ge \frac{(\vec{F}_{ext, i} - \vec{F}_{ext, i-1}) \cdot \hat{l}_i + 2F_{c, i} + \hat{l}_{i-1} \cdot \hat{l}_i T_{i-1} + \hat{l}_{i+1} \cdot \hat{l}_i T_{i+1}}{2} \\
     T_1  &\ge   \vec{F}_{ext, 1} \cdot \hat{l}_1 + F_{c, 1} - M_1 \ddot{\vec{x}}_0 \cdot \hat{l}_1 + \hat{l}_2 \cdot \hat{l}_1 T_2 
\end{align}
\end{subequations}
and enforce them sequentially (from top to bottom in Equation \ref{eqn:gaussseidel_set}) by choosing each $T_i$ equal to the larger between zero and right hand side.

After solving for $T_i$ via Equation \ref{eqn:gaussseidel_set}, Equation \ref{eqn:netforce_all} can be used to find the net force $\vec{F}_{net, i}$ on each non-kinematic virtual bone.
Given $\vec{F}_{net, i}$ for each non-kinematic virtual bone, the velocity can be updated in any order (for both step 1 and step 4 of the time integration);
in addition, the velocity of the kinematic virtual bone (the root) should be updated as well.

The updated velocities may be subject to an additional instantaneous impulse of magnitude $I_i \ge 0$ whenever a rope is at its maximal length $l_{max, i}$.
Let $\vec{v}_{pre, i}$ and $\vec{v}_{post, i}$ be the velocity before and after (respectively) this instantaneous exchange of momentum; then, the impulses along the ropes are applied via 
\begin{subequations}
\begin{align}
    \label{eqn:impulse}
    M_i \vec{v}_{post, i} &= M_i \vec{v}_{pre, i} - I_{i} \hat{l}_{i} + I_{i+1} \hat{l}_{i+1} \\
    \label{eqn:impulse_m}
    M_m \vec{v}_{post, m} &= M_m \vec{v}_{pre, m} - I_{m} \hat{l}_{m}
\end{align}
\end{subequations}
where $i \in \{1, ..., m-1\}$ and $I_i=0$ whenever $l_i < l_{max, i}$. Note that $\vec{v}_{post, 0} = \vec{v}_{pre, 0}$, since $M_0=\infty$. In order to prevent each taut rope from overstretching,
\begin{align}
    \label{eqn:impulseineq}
    (\vec{v}_{post, i} - \vec{v}_{post, i-1}) \cdot \hat{l}_i \le 0
\end{align}
must hold, where $i \in \{1, ..., m\}$ . Substituting Equations \ref{eqn:impulse} and \ref{eqn:impulse_m} into Equation \ref{eqn:impulseineq} gives
\begin{subequations}
\label{eqn:impulse_set}
\begin{align}
    -\frac{1}{M_1} I_1 + \frac{\hat{l}_{2} \cdot \hat{l}_1}{M_{1}} I_{2} \le (\vec{v}_{pre, 0}- \vec{v}_{pre, 1}) \cdot \hat{l}_1 \label{eqn:impulse_system_1} \\
    \frac{\hat{l}_{i-1} \cdot \hat{l}_i}{M_{i-1}} I_{i-1} - (\frac{1}{M_{i-1}} + \frac{1}{M_i}) I_i + \frac{\hat{l}_{i+1} \cdot \hat{l}_i}{M_{i}} I_{i+1} \le (\vec{v}_{pre, i-1}- \vec{v}_{pre, i}) \cdot \hat{l}_i \label{eqn:impulse_system} \\
    \frac{\hat{l}_{m-1} \cdot \hat{l}_m}{M_{m-1}} I_{m-1} - (\frac{1}{M_{m-1}} + \frac{1}{M_m}) I_m \le (\vec{v}_{pre, m-1}- \vec{v}_{pre, m}) \cdot \hat{l}_m\label{eqn:impulse_system_m}
\end{align}
\end{subequations}
where $i \in \{2, ..., m-1\}$.
This tri-diagonal linear system of inequalities decouples into separate blocks whenever a slack rope has $l_i < l_{max, i}$ or two consecutive ropes are orthogonal.
For computational efficiency, we iterate these equations in root-to-tip order (from top to bottom in Equation \ref{eqn:impulse_set}) using Gauss-Seidel; for most real-time applications, typically only one root-to-tip sweep is required.
In particular, we rewrite \ref{eqn:impulse_system_1}, \ref{eqn:impulse_system}, and \ref{eqn:impulse_system_m} as
\begin{subequations}
\label{eqn:impulse_gs_set}
\begin{align}
    I_1 &\ge M_1 \left( (\vec{v}_{pre, 1}- \vec{v}_{pre, 0}) \cdot \hat{l}_1 + \frac{\hat{l}_{2} \cdot \hat{l}_1}{M_{1}} I_{2} \right) \\
    I_i &\ge \frac{M_{i-1}M_i}{M_{i-1} + M_i} \left( (\vec{v}_{pre, i}- \vec{v}_{pre, i-1}) \cdot \hat{l}_i + \frac{\hat{l}_{i-1} \cdot \hat{l}_i}{M_{i-1}} I_{i-1} + \frac{\hat{l}_{i+1} \cdot \hat{l}_i}{M_{i}} I_{i+1} \right) \\
    I_m &\ge \frac{M_{m-1}M_m}{M_{m-1} + M_m} \left( (\vec{v}_{pre, m}- \vec{v}_{pre, m-1}) \cdot \hat{l}_m + \frac{\hat{l}_{m-1} \cdot \hat{l}_m}{M_{m-1}} I_{m-1} \right)
\end{align}
\end{subequations}
and enforce them sequentially (from top to bottom in Equation \ref{eqn:impulse_gs_set}) by choosing each $I_i$ equal to the larger between zero and right hand side.

\begin{figure}
    \includegraphics[width=\columnwidth]{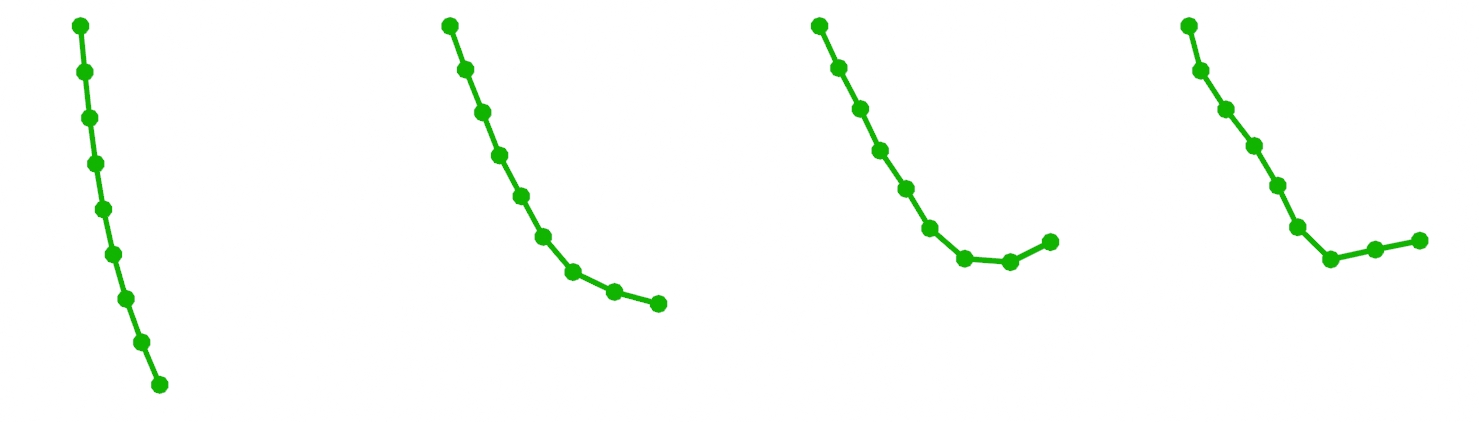}
  \caption{Motion of a single swinging rope chain, showcasing varying numbers of Gauss-Seidel iterations. Left to right: 1, 5, 10 iterations, and iterating until the relative error is smaller than a tolerance of $10^{-6}$. With more iterations, the rope chain is less damped (as expected). The same number of iterations (or the same tolerance) was used for both the tension and impulse computations. Note that it might be more efficient (depending on the application) to use a different number of iterations on the tension and impulse computations.}
  \label{fig:gauss_seidel}
\end{figure}

Note that we chose tip to root for the tension so that each virtual bone feels the weight of all the virtual bones below it even when executing only one iteration. Conversely, we chose root to tip for the impulse since it has been argued (and shown) to work well for contact resolution in prior works (see e.g. \cite{guendelman2003nonconvex}). Figure \ref{fig:gauss_seidel} illustrates the results one might typically expect with different numbers of iterations.

\section{Position Update}
\label{sec:posupdate}
In contrast to the tension and impulse computations discussed in Section \ref{sec:velupdate} which address force and velocity constraints respectively, a stricter approach is desirable for position constraints (especially to avoid errors in the rendered visualizations); 
thus, we execute one sweep from root to tip on the length of each rope in order to prevent it from exceeding its maximum length. 
Each virtual bone is updated from its time $t^n$ position to its time $t^{n+1}$ position by considering both its time $t^{n + \frac{1}{2}}$ velocity (computed as described in Section \ref{sec:velupdate}) and the rope constraint on its position relative to the previously updated virtual bone (in the root to tip sweep). That is, $\vec{x}_i^n$ is updated to $\vec{x}_i^{n+1}$ by considering both $\vec{v}_i^{n+\frac{1}{2}}$ and the rope that connects the virtual bone to $\vec{x}_{i-1}^{n+1}$.

When the rope is slack, the virtual bone's position can evolve freely with no hindrance from the constraint; however, when the rope is taut, the virtual bone is constrained to rotate on the sphere about $\vec{x}_{i-1}^{n+1}$.
Approximating this with an evolve-and-project strategy damps the rotation to the linearized rotation. 
This can be seen by projecting the distance moved in a linearized rotation 
\setlength{\intextsep}{1pt}%
\setlength{\columnsep}{5pt}%
\begin{wrapfigure}{r}{0.4\columnwidth}
\begin{center}
    \includegraphics[width=0.4\columnwidth]{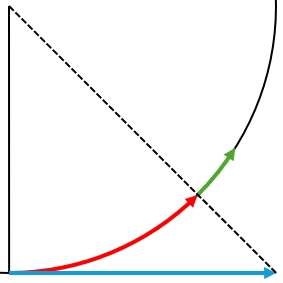}
\end{center}
\end{wrapfigure}
(colored blue in the figure) back onto a great circle and noting that the arc length thus traversed (colored red in the figure) is smaller than the distance covered in the linearized rotation, which is the arc length distance that should have been traversed (the sum of red and green colored arcs in the figure).  
Thus, we address the constrained motion with a non-linearized (actual) rotation.

The tautness of the rope is governed by the quadratic function
\begin{equation}
    f(s) = |\vec{x}^n_{i} + s \vec{v}^{n+\frac{1}{2}}_{i} - \vec{x}_{i-1}^{n+1}|^2 - l_{max, i}^2
\end{equation} 
where $f(s) < 0$ indicates slack and $f(s) >0$ indicates overstretching. When $f(\Delta t) \le 0$, the rope is not overstretched at the end of the time step and $\vec{x}^{n+1}_{i} = \vec{x}^n_{i} + \Delta t \vec{v}^{n+\frac{1}{2}}_{i}$ is accepted as the final position. Otherwise, when $f(\Delta t) > 0$, we compute the largest root $s_{root}$ in the interval $[0, \Delta t]$. If there is no root in the interval, then we set $s_{root} = 0$ and project the overstretched $\vec{x}^{n}_{i}$ back onto the surface of the sphere via
\begin{equation}
\vec{x}_{i}^{n} \leftarrow \vec{x}_{i-1}^{n+1} + \frac{\vec{x}_{i}^{n} - \vec{x}_{i-1}^{n+1}}{|\vec{x}_{i}^{n} - \vec{x}_{i-1}^{n+1}|}l_{max, i}
\end{equation}
so that it is no longer overstretched. In the interval $[0, s_{root}]$, the virtual bone is allowed to move unhindered by the constraint via $\vec{x}^{s_{root}}_{i} = \vec{x}^n_{i} + s_{root} \vec{v}^{n+\frac{1}{2}}_{i}$. Note that this intentionally ignores any overstretching in $[0, s_{root}]$, since such overstretching is overcome automatically and may only be the (spurious) result of updating the virtual bones one at a time (from root to tip) as opposed to uniformly. 

In the interval $[s_{root}, \Delta t]$, the virtual bone is constrained to rotate on the sphere centered at $\vec{x}_{i-1}^{n+1}$ of radius $l_{max, i}$.
Given $\vec{l}_{i}^{s_{root}} = \vec{x}_{i}^{s_{root}} - \vec{x}_{i-1}^{n+1}$ and $\hat{l}_{i}^{s_{root}} = \frac{\vec{l}_{i}^{s_{root}}}{|\vec{l}_{i}^{s_{root}}|}$, the tangential velocity 
\begin{equation}
    \vec{v}_{T} = \vec{v}^{n+\frac{1}{2}}_{i} - (\vec{v}^{n+\frac{1}{2}}_{i} \cdot \hat{l}_i^{s_{root}})\hat{l}_i^{s_{root}}
\end{equation}
is used to determine the distance $d_T = |\vec{v}_{T}|(\Delta t - s_{root})$ the virtual bone rotates on the great circle specified by the direction $\hat{v}_{T} = \frac{\vec{v}_{T}}{|\vec{v}_{T}|}$.
A rotation matrix $R$ is defined to rotate the virtual bone by an amount $\theta=\frac{d_T}{l_{max, i}}$ about the axis $\hat{l}_{\theta}=\hat{l}_i^{s_{root}} \times \hat{v}_T$ via 
\begin{align}
    \vec{l}^{n+1}_i = R (\theta, \hat{l}_{\theta}) \vec{l}_i^{s_{root}} 
\end{align}
to obtain $\vec{x}_{i}^{n+1}=\vec{x}_{i-1}^{n+1} + \vec{l}^{n+1}_i$.

After the position update, the velocities may no longer satisfy the constraint preventing overstretching (see Equation \ref{eqn:impulseineq}). Thus, new impulses can be computed and applied (following the discussion in the second half of Section \ref{sec:velupdate}). 
Alternatively, impulses can instead be computed and applied after resolving collisions, both after the position update and after resolving collisions, or deferred entirely (and left unmodified until the end of second velocity update).

\section{Collisions}
\label{sec:collision}
\begin{figure}
    \includegraphics[width=\columnwidth]{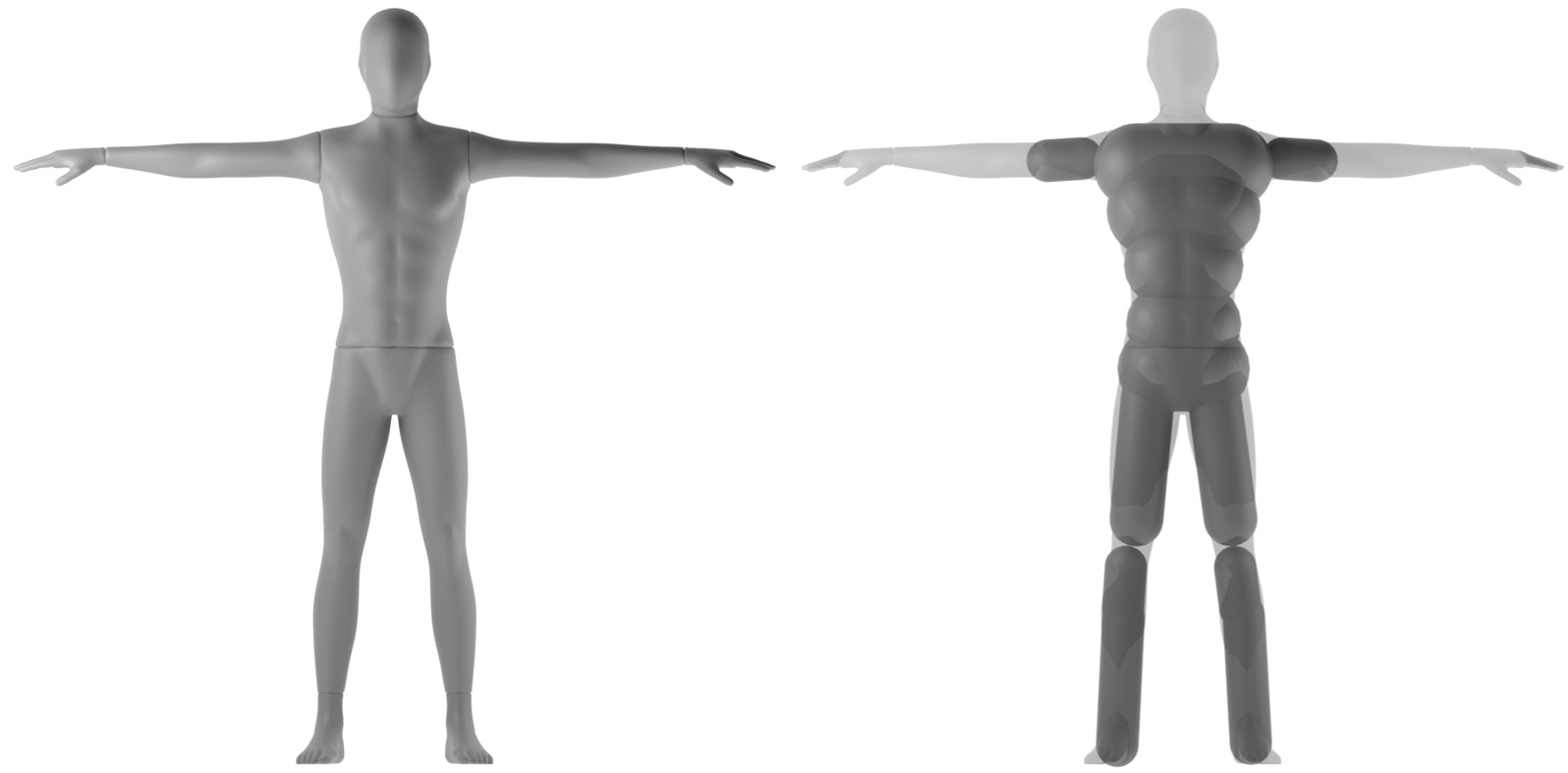}
  \caption{For the sake of computational efficiency, we represent the volumetric body (left) with a number of analytically defined SDFs (right). Although other representations (three dimensional grid discretizations, neural representations, etc.) are compatible with our approach, they incur higher computational costs.}
  \label{fig:sdf}
\end{figure}

Generally speaking, we follow the method in \cite{bridson2003simulation} detecting and processing collisions between each non-kinematic virtual bone and each signed distance function (SDF). Although this approach has quite efficient $O(1)$ processing time for each virtual bone, it is computationally expensive to load three dimensional SDF discretizations into memory; in addition, the limited memory availability in most real-time systems makes it infeasible to store the SDFs persistently. Thus, we avoid these costly three dimensional discretizations by utilizing SDFs that can be defined analytically (see e.g.  \cite{blinn1982generalization, wyvill1986soft, nishimura1985object}). See Figure \ref{fig:sdf}. Alternatively, a number of authors have aimed to represent SDFs via neural networks (see e.g. \cite{park2019deepsdf, remelli2020meshsdf}). If the number of parameters in the neural network can be made to be much smaller than an equivalently accurate three dimensional discretization, then the computational costs associated with high demands on memory could be avoided. Instead of aiming to make each neural SDF utilize less memory than its equivalently accurate three dimensional discretization, one could aim to minimize the additional memory burden incurred by switching from one neural SDF to another (e.g. by using shape descriptors). 

Motivated by the cloth-object collision discussion in \cite{selle2008robust}, we propose a modification to \cite{bridson2003simulation} in order to obtain more robust behavior for real-time applications with coarse discretizations and large time steps. 
In the standard approach, an initially non-interpenetrating particle that ends up in the interior of an SDF is pushed outwards in the $\nabla \phi$ direction until it reaches the surface of the SDF (or a bit further than the surface when aiming for wrinkle preservation as discussed in \cite{bridson2003simulation}).
In contrast, a real-world particle is unable to penetrate into the interior of an object and instead collides with the surface and responds accordingly.
In the limit as the time step goes to zero and the number of points used to represent the surface goes to infinity, the standard numerical approach converges to the correct real-world solution (roughly speaking, ignoring the inability to properly model friction, microscopic structure, etc.).
However, the numerical errors are exacerbated by the large time steps and the coarse surface discretizations utilized for real-time applications.
Although continuous collision detection (CCD) could be used to increase the accuracy while maintaining a large time step, CCD is too computationally burdensome for most real-time applications (although progress is being made, see e.g. \cite{li2020p} and the references therein).

The standard approach of evolving a particle into an interpenetrating state and subsequently pushing it outwards in the $\nabla \phi$ direction can be thought of as a predictor-corrector method modeling the actual path of the particle (the path that CCD would aim to trace out). Aiming to preserve the computational efficiency of the predictor, we propose modifying the corrector in order to obtain a more accurate final state. This can be done efficiently by choosing a more appropriate direction for push out than $\nabla \phi$. In fact, $\nabla \phi$ can be highly erroneous when objects are thin (causing a particle to be pushed to the wrong side) or have high curvature (causing a particle to be pushed to the wrong direction); moreover, these errors are exacerbated by the larger time steps typically used in real-time applications. Given limited information, our ansatz is that the safest push out direction is the reverse direction along the path the particle traversed as it penetrated into the collision body. At the very least, this aims to return the particle to the point where a CCD collision would have occurred.

Assuming the collision body is stationary, the reverse path out of the collision body back towards the CCD collision point has direction $\vec{r} = \vec{x}^{n} - \vec{x}^{n+1}$ where $\vec{x}^{n+1}$ is the predicted position of the particle (penetrating into the collision body). The main difficulty associated with using this direction is that it is unclear how far the particle should move. There are several options for addressing this. One could use the local value of $|\phi|$ as usual, but this does not necessarily alleviate interpenetration when $\hat{r}=\frac{\vec{r}}{|\vec{r}|}$ and $\nabla \phi$ point in different directions. Notably, the local value of $|\phi|$ is always too small, except when $\hat{r} = \nabla \phi$ in which case the particle should exactly reach the surface of the collision body. Therefore, one could iterate the push out a few times in order to better approach the surface. In addition, one could use a small $\epsilon > 0$ to augment the local value of $|\phi|$. This is roughly equivalent to using an SDF thickened by $\epsilon$ or to pushing the particle outwards to the $\phi=\epsilon$ isocontour. Finally, note that one could use line search (perhaps via bisection), which is equivalent to CCD for the simple case when the collision body is stationary (CCD is typically much cheaper in this simple case).

When the collision body is moving, $\vec{x}^n$ is not necessarily non-interpenetrating (since the collision body may move to cover it); in such a scenario, $\hat{r}$ is no longer guaranteed to be a suitable replacement for $\nabla \phi$. Fortunately, this is easily remedied by analyzing the problem in the moving frame of the collision body. 
As the collision body moves and deforms from time $t^n$ to $t^{n+1}$, we embed the particle to move with it. This guarantees that the new particle location, denoted $\vec{x}_B^n$, is non-interpenetrating. For example, when the collision body moves rigidly, that same rigid motion is applied to $\vec{x}^n$ to obtain $\vec{x}_B^n$. For deforming and/or skinned collision bodies, the deformation/skinning needs to be extended to $\vec{x}^n$ in order to obtain $\vec{x}_B^n$. This allows $\vec{r} = \vec{x}_B^n - \vec{x}^{n+1}$ to be used as a valid push out direction.

\begin{figure*}
    \includegraphics[width=\textwidth]{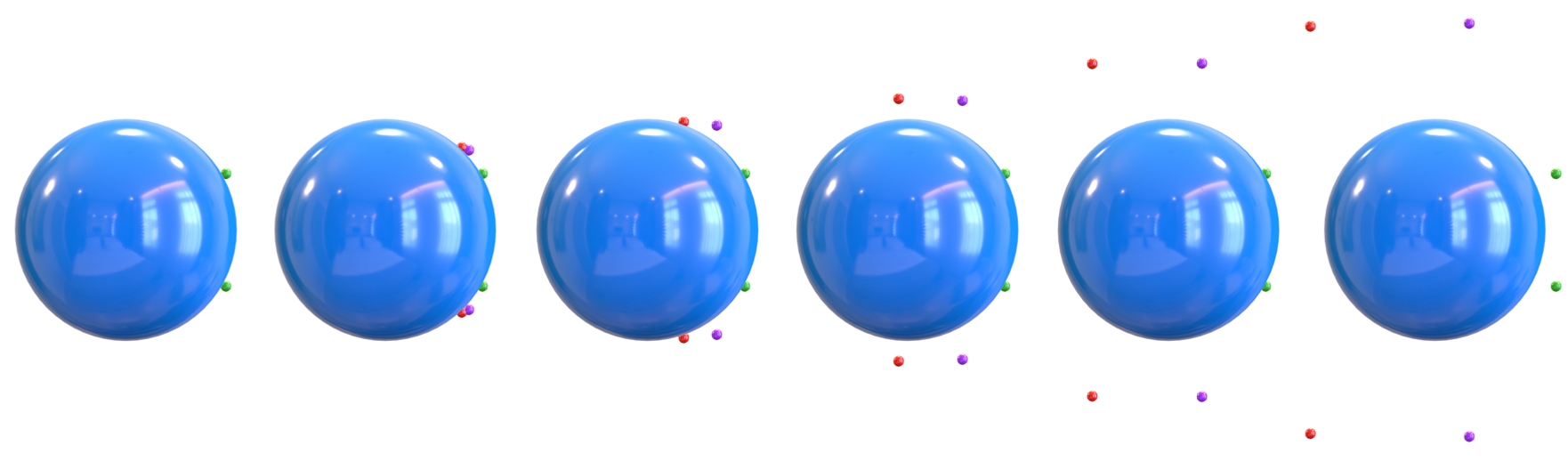}
    \caption{Time evolution of a sphere colliding with six particles (6 frames are depicted). The red particles use the $\nabla \phi$ direction for both push out and velocity projection (as is typical), the purple particles use $\hat{r}$ for push out and $\nabla \phi$ for velocity projection, and the green particles use $\hat{r}$ for both push out and velocity projection. Note how replacing $\nabla \phi$ with $\hat{r}$ prevents the particles from quickly working their way around the sphere. In fact, the green particles maintain a persistent contact with the sphere until it stops moving shortly before the last frame (even though friction is not used in this example). The behavior of the green particles is highly preferable to that of the red (and purple) particles when considering collisions between clothing and the human body.}
    \label{fig:toy_coll}
\end{figure*}

After the position $\vec{x}^{n+1}$ is modified to be collision-free, the velocity $\vec{v}^{n+\frac{1}{2}}$ is modified to ensure that the relative normal velocity $(\vec{v}^{n+\frac{1}{2}} - \vec{v}_{\phi}) \cdot \hat{N}$ does not point into the SDF via 
\begin{equation}
    v^{new}_N = \max(\vec{v}^{n+\frac{1}{2}} \cdot \hat{N} , \vec{v}_{\phi} \cdot \hat{N})
\end{equation}
where $\vec{v}_{\phi}$ is the velocity of the (extended, if necessary) collision body at $\vec{x}^{n+1}$, and $\hat{N}$ may be chosen as either $\nabla \phi$ or $\hat{r}$. The relative tangential velocity
\begin{equation}
    \vec{v}_{rel, T} = \vec{v}^{n+\frac{1}{2}}   - \vec{v}^{n+\frac{1}{2}} \cdot \hat{N} -  \vec{v}_{\phi, T}
\end{equation}
is defined using the tangential velocity $\vec{v}_{\phi, T} = \vec{v}_{\phi} - \vec{v}_{\phi} \cdot \hat{N}$ of the (extended, if necessary) collision body at $\vec{x}^{n+1}$. When the friction coefficient $\mu$ is non-zero, the relative tangential velocity is modified via
\begin{align}
    \vec{v}^{new}_T &= \vec{v}_{\phi, T} + \max(0, 1 - \mu \frac{v^{new}_N - \vec{v}^{n+\frac{1}{2}} \cdot \hat{N}}{|\vec{v}_{rel, T}|} )\vec{v}_{rel, T}
\end{align}
as suggested by \cite{bridson2002robust}. The final post collision velocity is given by $v_N^{new}\hat{N} + \vec{v}_T^{new}$.
See Figure \ref{fig:toy_coll}.

The collisions are processed sequentially from root to tip. 
Since collisions alter the positions of the virtual bones, length constraints are enforced in this step as well. This is accomplished by adjusting the position of a virtual bone via $\vec{x}_i^{new}=\vec{x}_{i-1} + l_{max, i} \hat{l}_i$ whenever $l_i > l_{max, i}$. Of course, this can create new collisions, so back-and-forth iteration is desirable. The root to tip sweep is done only once, and any back-and-forth iteration between the collision and the length constraint happens only once for each virtual bone. We recommend starting this iteration with the length constraint, since it may remove the need for collisions. We also stress the importance of finishing this iteration with the collision check in order to preserve a non-interpenetrating state. 

\section{Neural Skinning}
\label{sec:skinning}

After each frame of rope chain simulation, a full cloth mesh needs to be reconstructed for rendering. After extensive experimentation, we were not able to obtain reasonable results via any of the standard skinning methods (such as LBS \cite{magnenat1988joint} or Dual Quaternion Skinning \cite{kavan2007skinning}). We were also unable to remedy these issues with a corrective quasistatic neural network (see Section \ref{sec:qnn}). The difficulties are likely due to the rope chain simulation's inability to produce good rotational information for the virtual bones, even with various procedural modifications. Thus, we took an alternative approach that utilizes a neural network to infer PCA coefficients for the cloth mesh from the virtual bone translational degrees of freedom (only). This resulted in a mesh suitable enough for a corrective QNN to operate on (see Section \ref{sec:qnn}). 

For each cloth mesh under consideration, we utilize on the order of 5000 frames of simulated data (any offline simulation system will do) in order to construct a standard PCA model based on non-rigid displacements from the cloth rest state. The rest state of the cloth mesh is defined by its steady-state draped position in the rest pose of the body. Given an animated body pose, the rigid component of the cloth displacement is calculated as the rigid displacement of a key body part (the neck for the cape and the pelvis for the skirt) and removed from the cloth displacement in order to obtain its non-rigid displacement. About 100 PCA basis functions are retained for neural skinning. 
When the cloth meshes were converted to rope chains (see Figure \ref{fig:pendsetup}), not all virtual bones were simulated (see the non-interconnected virtual bones in Figure \ref{fig:pendsetup}); however, these virtual bones are needed in order to reconstruct the full cloth mesh and as such are also used as input into the neural skinning network.
Given non-rigid displacements of the virtual bones from their positions in the cloth mesh rest state (their rigid component is identical to that used for the cloth mesh), the neural skinning network is trained to infer the approximately 100 PCA coefficients used to reconstruct the full cloth mesh.

We utilize a lightweight 2-layer MLP with 500 neurons per layer. In order to train the network, the same 5000 frames of simulation data (previously described) is used for supervision. The virtual bones are embedded in the cloth rest state and barycentrically enslaved to move with the simulation data yielding the inputs for the network. For each frame utilized in the data term of the loss function, the network-inferred PCA coefficients are used to reconstruct a cloth mesh that is compared to the simulation ground truth vertex-by-vertex via a standard L2 norm. Importantly, each frame is treated as non-sequential in order to reduce the need for a higher capacity neural network, minimize the burden associated with collecting more training data, and avoid unnecessary overfitting. We did not find the need to add any additional terms (such as Laplacians) that would regularize the cloth mesh, since the use of a PCA model already provides sufficient regularization.

Even when the ground truth cloth meshes used in training do not interpenetrate the body, the inferred results will typically contain interpenetrations. This is caused by the regularization used to combat overfitting and to increase the efficacy of generalization to unseen data. In order to alleviate interpenetration without adversely affecting desirable regularization, we include a PINN-style collision loss during training. When a cloth vertex is found to be interpenetrating, we calculate a suitable non-interpenetrating position for that vertex (using push out) and include the difference between the vertex and its non-interpenetrating state in the PINN-style collision loss. The non-interpenetrating state can be calculated using $\nabla \phi$ as the push out direction along with the $\phi$ value (with or without iteration) or CCD to calculate the push out distance. Given the potentially large inaccuracy of the cloth state during training, $\nabla \phi$ often points in an unhelpful direction; thus, we instead chose an alternative push out direction $\hat{r}$ pointing from the current position to the ground truth position. Note that it can be desirable for the non-interpenetrating state to be well-separated from the collision body (not just on the zero-isocontour) in order to create a buffer on the vertices that helps to alleviate edge interpenetrations (and/or to preserve wrinkling, see e.g. \cite{bridson2003simulation}). Since we detach the non-interpenetrating state from the automatic differentiation graph, an interpenetrating vertex's contribution to the gradient from the PINN-style collision loss is parallel to its contribution from the data term. This can be interpreted as increasing the importance of matching the ground truth for vertices that are interpenetrating as compared to vertices that are non-interpenetrating; in fact, we use a weight of 1000 on the PINN-style collision loss and a weight of only 0.1 for the data term.

It is important to note that there is a large discrepancy between the virtual bone configurations that arise from rope chain simulations and the configurations in the training data, which are obtained by barycentrically embedding virtual bones in a mass-spring cloth simulation mesh. That is, we are aiming to make the network generalize well to (sometimes significantly) out-of-distribution data. The usual process of utilizing holdout data does help to increase the ability of a network to generalize to unseen data; however, the errors can still be quite large when the training data and holdout data come from one distribution and the unseen data comes from another. This is more akin to a domain gap. Both the PINN collision loss and the regularization obtained via the PCA model help to alleviate these issues with out-of-distribution unseen data. Even though the network-inferenced cloth may not follow the dynamic trajectory of the virtual bones as closely as one might prefer, the cloth tends to have an aesthetically pleasing shape and be interpenetration-free. 

\section{Neural Shape Inference}
\label{sec:qnn}
\begin{figure}
    \includegraphics[width=\columnwidth]{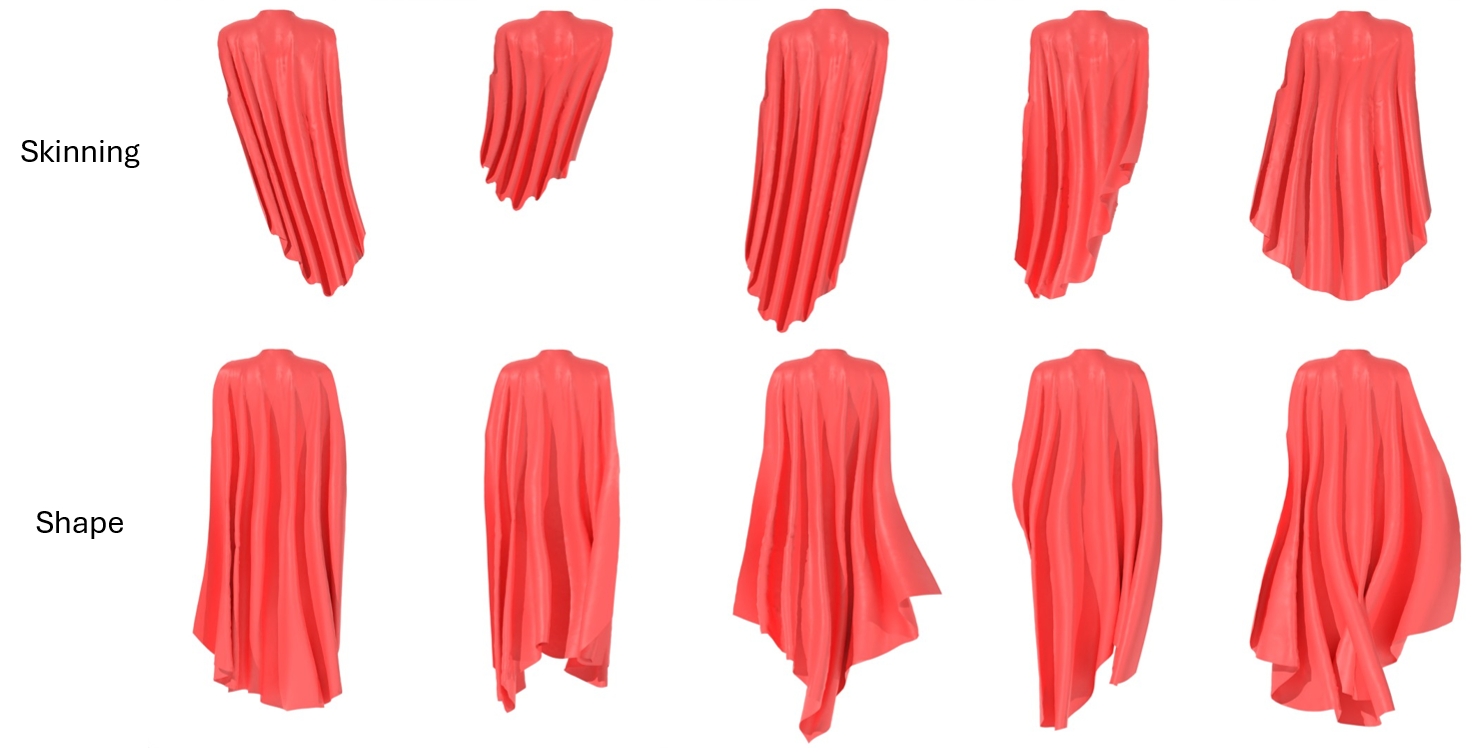}
    \caption{Top row: first five principle components of the neural skinning PCA model for the cape. Bottom row: first five principle components of the neural shape PCA model for the cape. All of the images depict the augmentation of the rest shape cloth mesh by the PCA displacements, even though the displacements are added to the skinned mesh (not the rest state mesh) during neural shape inference. The PCA model used for skinning tends to capture low-frequency deformations, while the PCA model used for shape inference is better suited for capturing higher-frequency deformations.}
    \label{fig:pca}
\end{figure}

Similar to the neural skinning in Section \ref{sec:skinning}, we utilize a 2-layer MLP with 500 neurons per layer in order to inference about 100 PCA coefficients from the non-rigid displacements of the virtual bones from their positions in the cloth rest state. The same 5000 frames of simulated data is used for training; however, the PCA model is calculated based on the non-rigid displacement between the output of the neural skinning network and the ground truth (see \cite{wang2024multi} for interesting discussion on multi-stage approaches). While the neural skinning is responsible for constructing a coarse approximation to the cloth mesh, the neural shape inference focuses on capturing more of the high-frequency spatial detail. This is obvious when comparing the two PCA models (See Figure \ref{fig:pca}). Similar to Section \ref{sec:skinning}, a PINN-style collision loss is also included.

\section{Results and Discussion}
\label{sec:results}

For both the neural skinning and the neural shape networks, we used a weight of 0.1 on the data term and weight of 1000 on the PINN-style collision loss. The initial learning rate for Adam \cite{kingma2014adam} was set to $10^{-4}$ to train the neural skinning network and $10^{-5}$ to train the neural shape network. A smaller learning rate was specified for the neural shape network, since its PCA coefficients represent smaller displacements. A cosine annealing schedule was used to decay the learning rate over epochs. The collision body SDFs were expanded by a small amount, and only one iteration was used for push out (for computational efficiency).

\begin{figure}
    \includegraphics[width=\columnwidth]{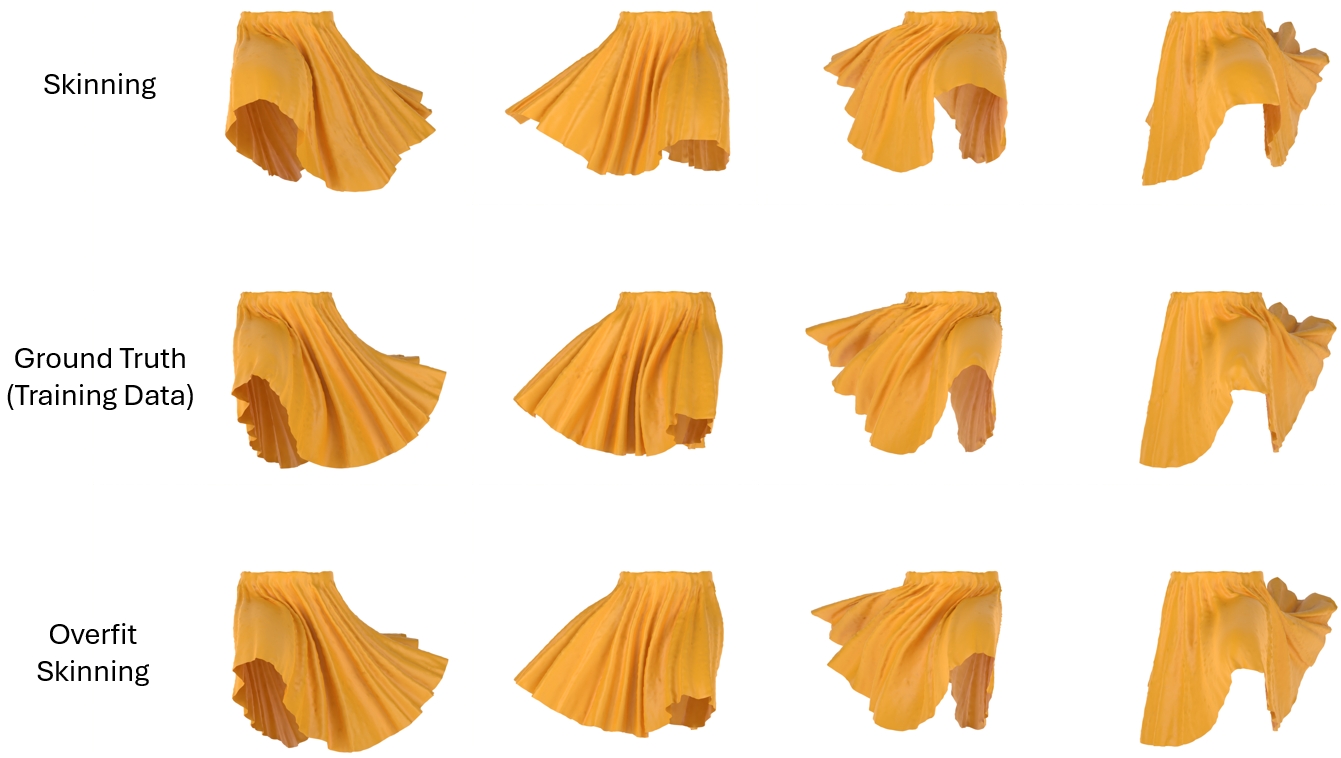}
    \caption{The first row shows the results of the neural skinning network, which can be compared to the ground truth training data in the second row. In order to demonstrate that the network does have the ability to match the ground truth data, the third row shows the overfit results obtained via overtraining; of course, an overfit network will not generalize well to unseen data.}
    \label{fig:train_skinning}
\end{figure}

\begin{figure}
    \includegraphics[width=\columnwidth]{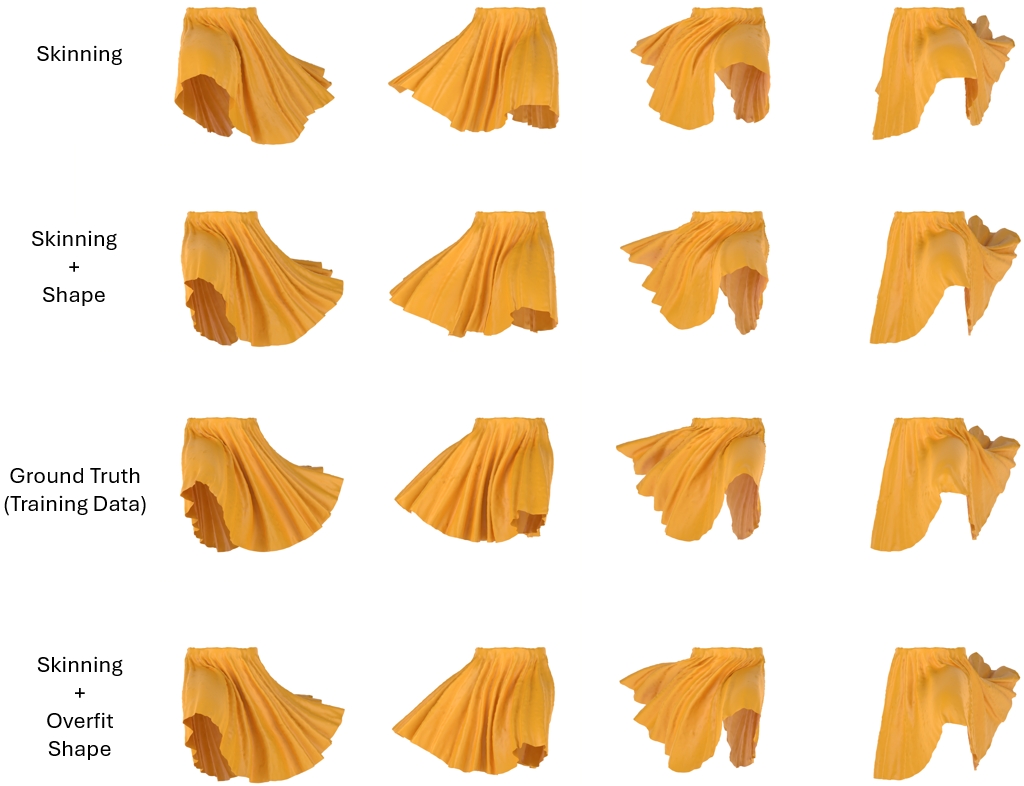}
    \caption{The first row shows the results of the neural skinning network (identical to the first row in Figure \ref{fig:train_skinning}). The second row shows the results of the neural shape network applied on top of the neural skinning result. The third row shows the ground truth training data. The fourth row shows how overtraining the neural shape network leads to results that well match (albeit overfit, similar to the last row in Figure \ref{fig:train_skinning}) the ground truth training data.}
    \label{fig:train_qnn}
\end{figure}

\begin{figure}
    \includegraphics[width=\columnwidth]{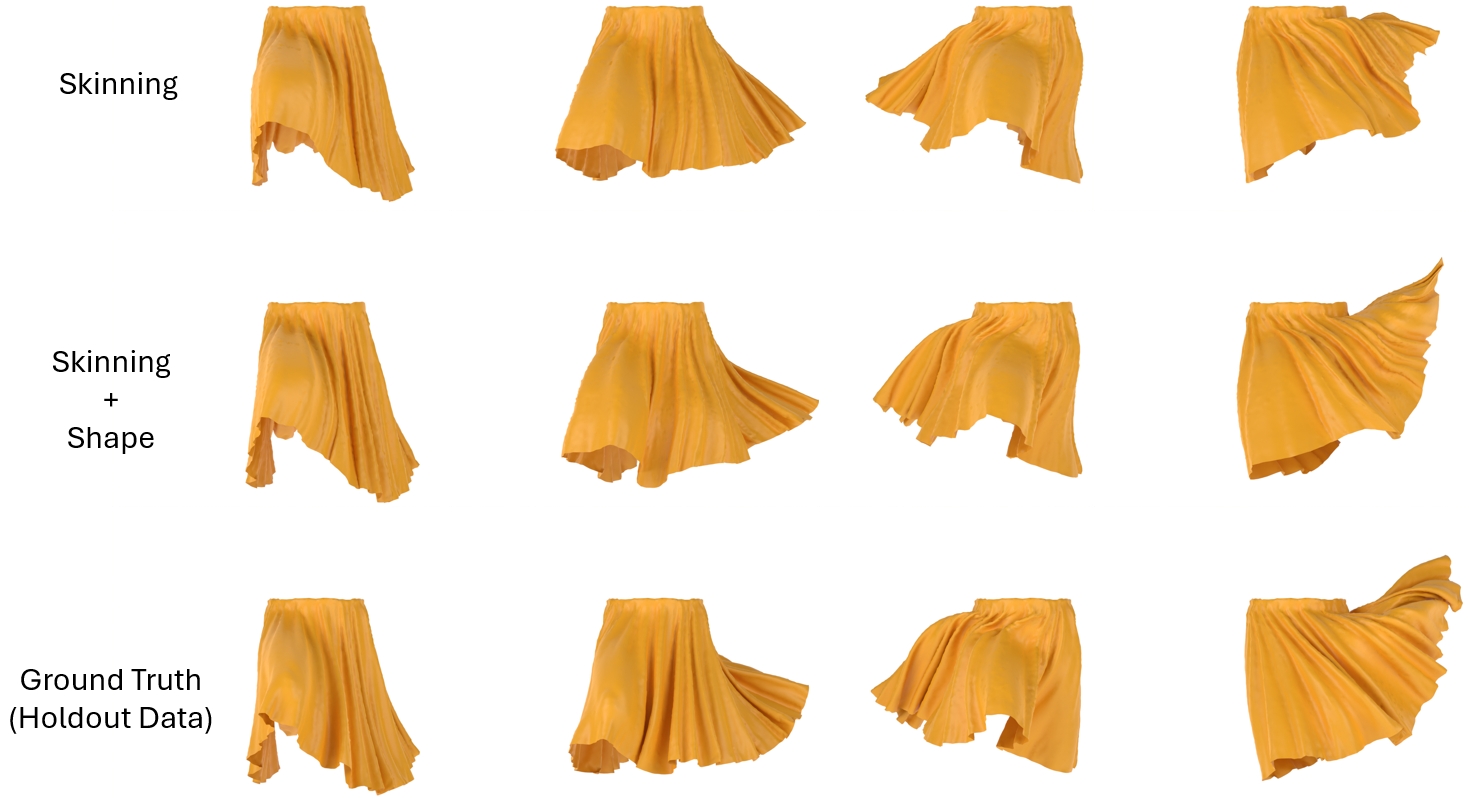}
    \caption{The first row shows the results of the neural skinning network, and the second row shows the results of the neural shape network applied on top of the neural skinning result. These holdout frames from the training set give some indication of how the network will perform on unseen data. Note that the network needs to generalize to out-of-distribution data (from rope chain simulations, as is discussed in the last paragraph in Section \ref{sec:skinning}), not just to holdout frames from the training set.}
    \label{fig:train_holdout}
\end{figure}

\begin{figure}
    \includegraphics[width=\columnwidth]{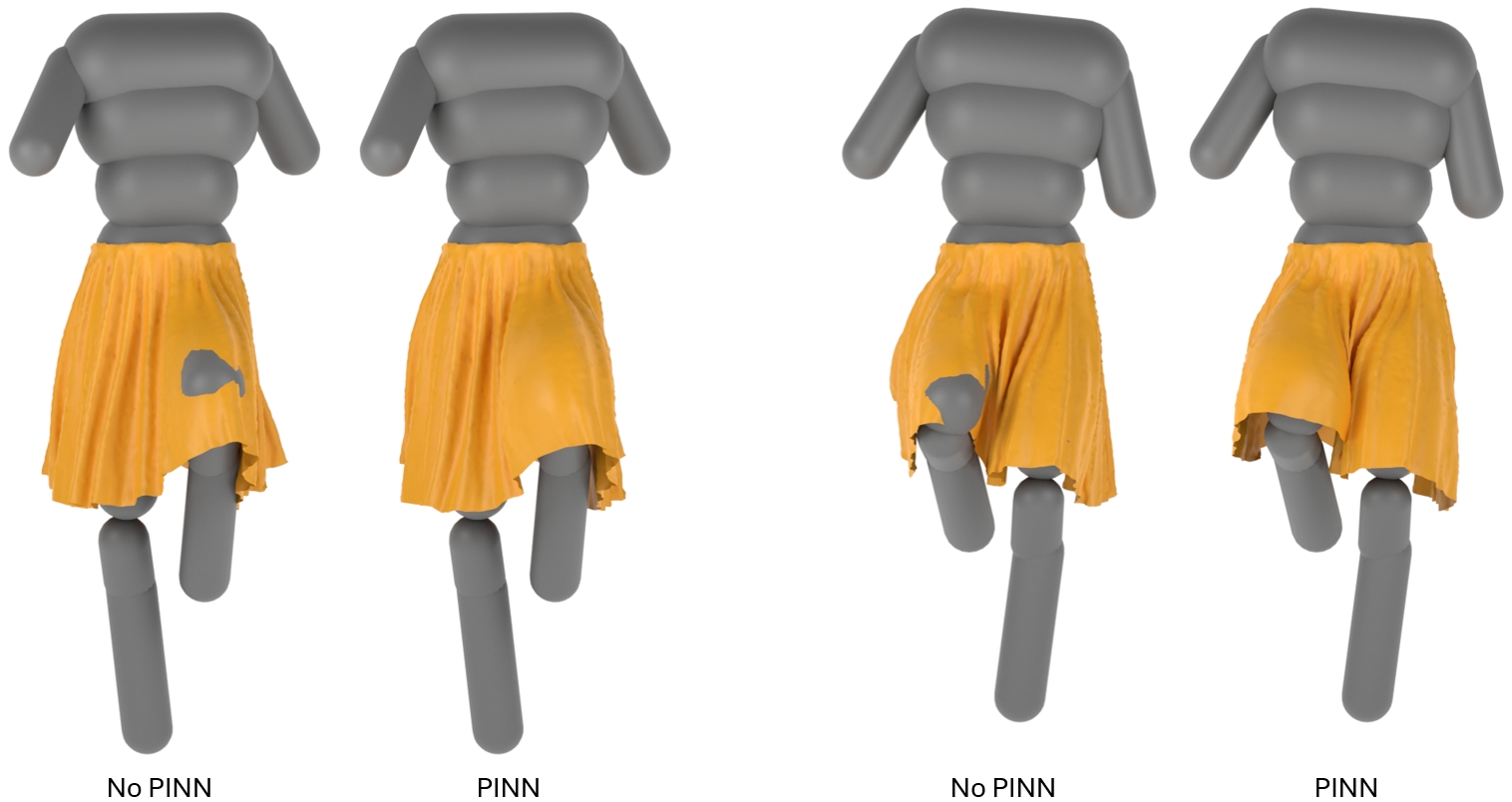}
    \caption{Comparing the result of a neural skinning network trained without and with the PINN-style collision loss.}
    \label{fig:pinn}
\end{figure}

We obtained our 5000 frames of training data (for both the cape and the skirt) using Houdini Vellum. 
80 percent of the frames were used in the loss function to train the neural network, and 10 percent of the frames (unseen in training) were used in a validation loss in order to choose parameters that might generalize well.
Figures \ref{fig:train_skinning} and \ref{fig:train_qnn} demonstrate the efficacy of the neural skinning network and the neural shape network respectively.
Another 10 percent of the frames were kept as true holdout data, in order to predict the ability of the networks to generalize to unseen (but still in-distribution) data. See Figure \ref{fig:train_holdout}. Finally, Figure \ref{fig:pinn} showcases the efficacy of the PINN-style collision loss.

\begin{figure*}
    \includegraphics[width=\textwidth]{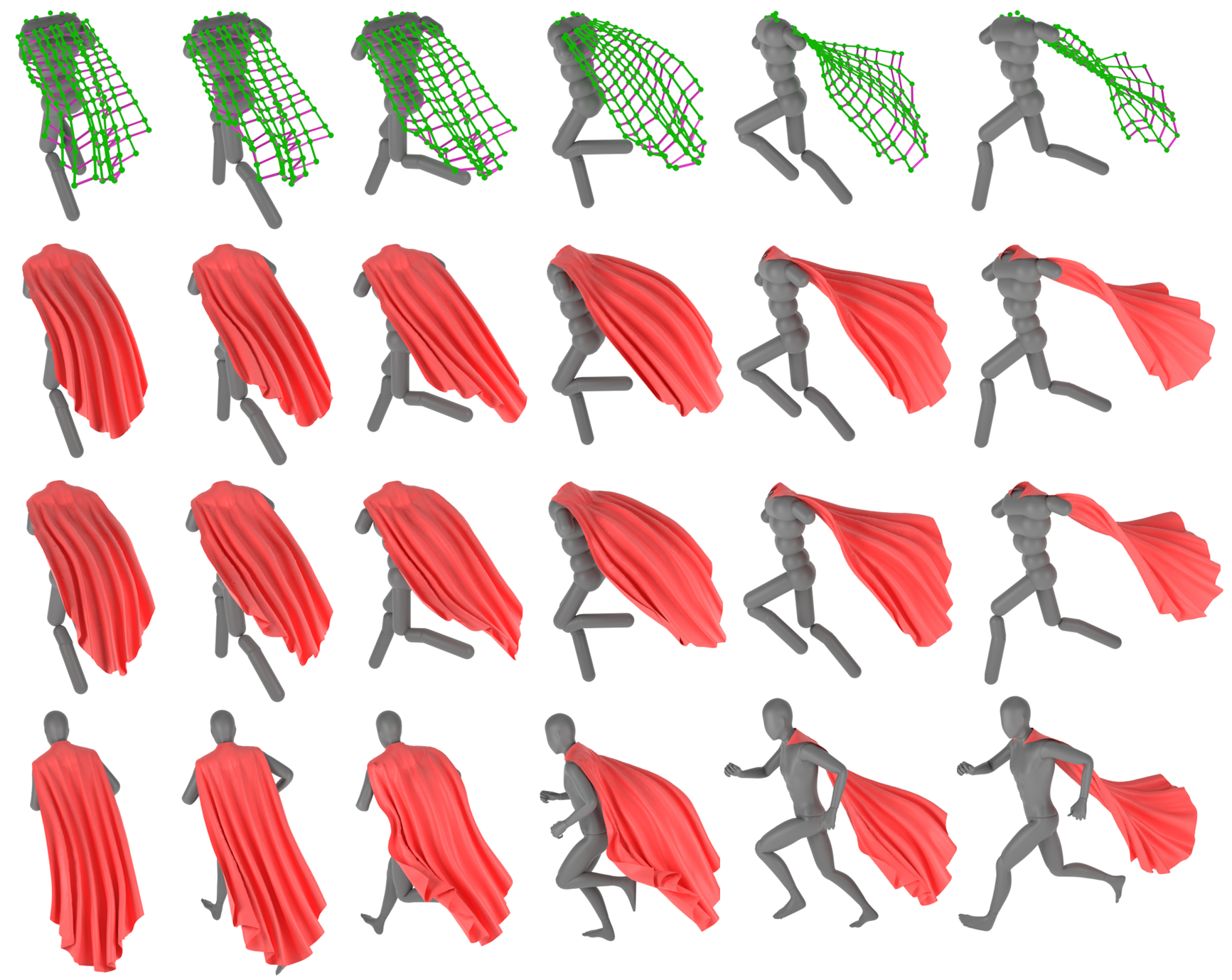}
    \caption{Simulation of a cape on an animation of a running person. First row: rope chain simulation. The rope chains are shaded green, and the weak lateral springs are visualized as purple edges. Second row: neural skinning, based on the first row. Third row: neural shape inference, based on the second row. Fourth row: Houdini cloth simulation with approximately 12K vertices (as a reference). The Houdini simulation and our rope chain approach both produce good dynamics, although  the Houdini simulation does exhibit visually displeasing erroneous over-stretching artifacts; in addition, the Houdini simulation is an order of magnitude slower than our currently unoptimized code.}
    \label{fig:cape_grid}
\end{figure*}

\begin{figure*}
    \includegraphics[width=\textwidth]{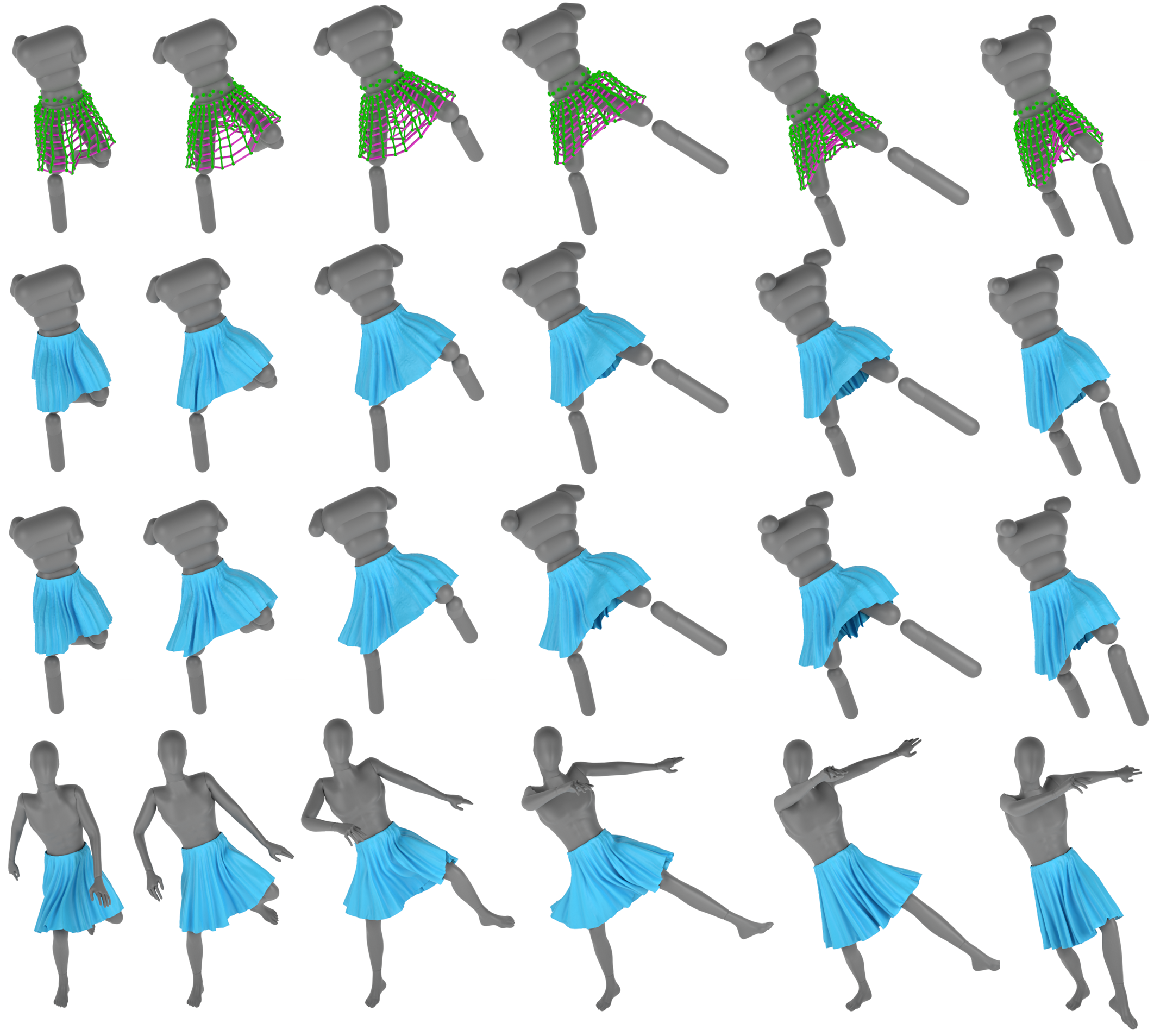}
    \caption{Simulation of a skirt on an animation of a dancing person. First row: rope chain simulation. The rope chains are shaded green, and the weak lateral springs are visualized as purple edges. Second row: neural skinning, based on the first row. Third row: neural shape inference, based on the second row. Fourth row: Houdini cloth simulation with approximately 18K vertices (as a reference). The skirt is rendered with a blue color in order to differentiate it from the yellow skirt figures, which use barycentrically embedded virtual bone positions instead of rope chain simulations as the network input.}
    \label{fig:skirt_grid}
\end{figure*}

In order to demonstrate how our networks generalize to unseen out-of distribution data from rope chain simulations, we considered two types of loose-fitting garments: capes and skirts. See Figures \ref{fig:cape_grid} and \ref{fig:skirt_grid}. Weak lateral springs were added in order to connect virtual bones from different rope chains (for both the skirt and the cape). Wind drag was added to the cape virtual bones. Damping, relative to the parent virtual bone in the chain, was added to the skirt virtual bones. The rope chains were collided against the analytic SDFs depicted in Figure \ref{fig:sdf}. Given the virtual bone positions obtained via rope chain simulation, we reconstructed the full cloth mesh using the neural skinning and neural shape networks. Even though the virtual bone positions generated from the rope chain simulations are out of distribution as compared to the training, validation, and holdout data, Figures \ref{fig:cape_grid} and \ref{fig:skirt_grid} demonstrate that the networks generalized well and obtained good results.

\begin{figure}
    \includegraphics[width=\columnwidth]{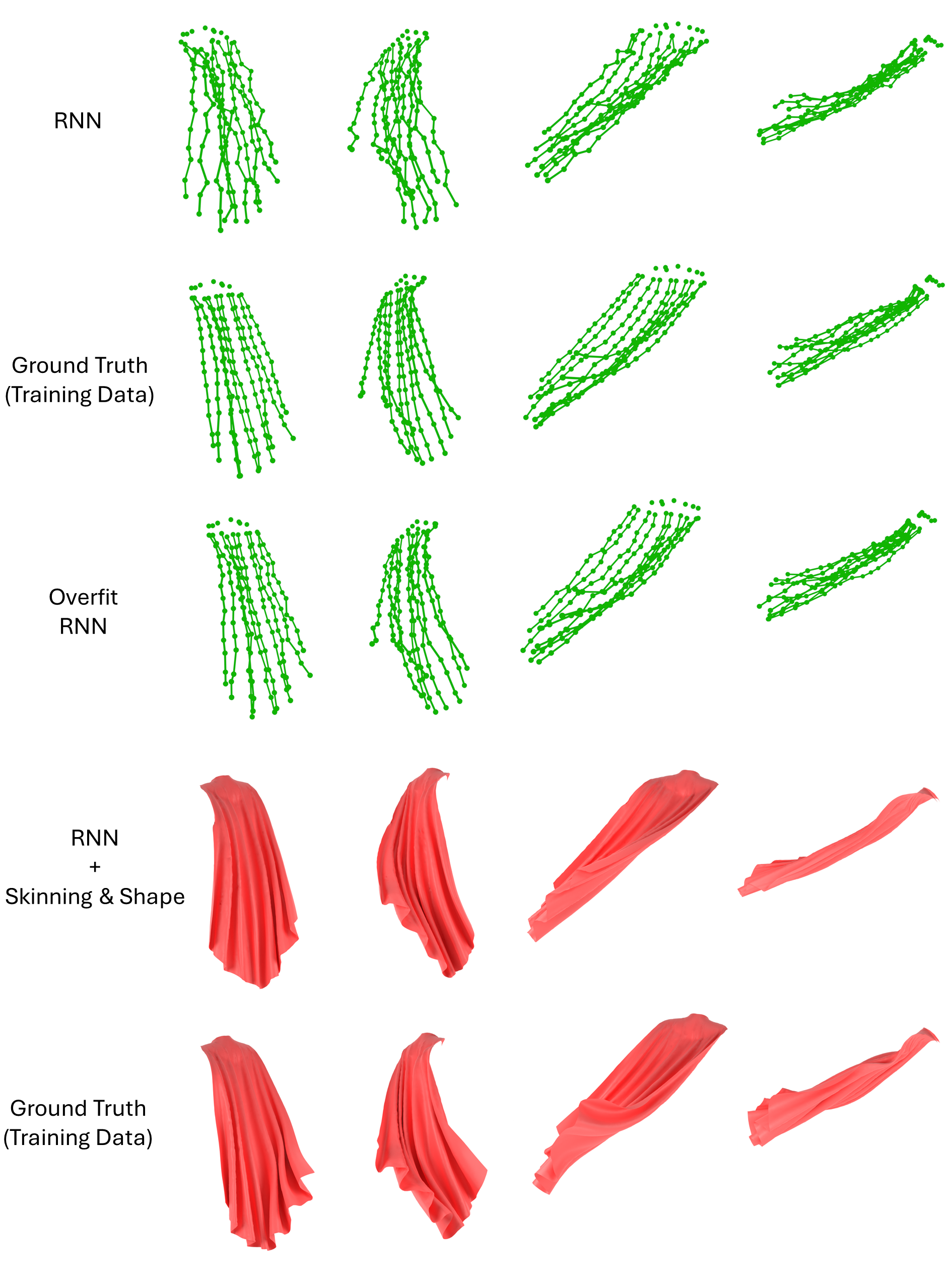}
    \caption{We trained a separate RNN for each rope chain (10 RNNs in total). The RNN results (first row) are quite noisy compared to the ground truth training data (second row). The thrid row shows the overfit results obtained via overtraining. The fourth row shows the result of applying our neural skinning and neural shape inference to the RNN-inferred virtual bone positions (from the first row). Comparing these results to the ground truth training data (fifth row) illustrates that our networks generalize well to this noisy RNN input.}
    \label{fig:rnn_training}
\end{figure}

\begin{figure}
    \includegraphics[width=\columnwidth]{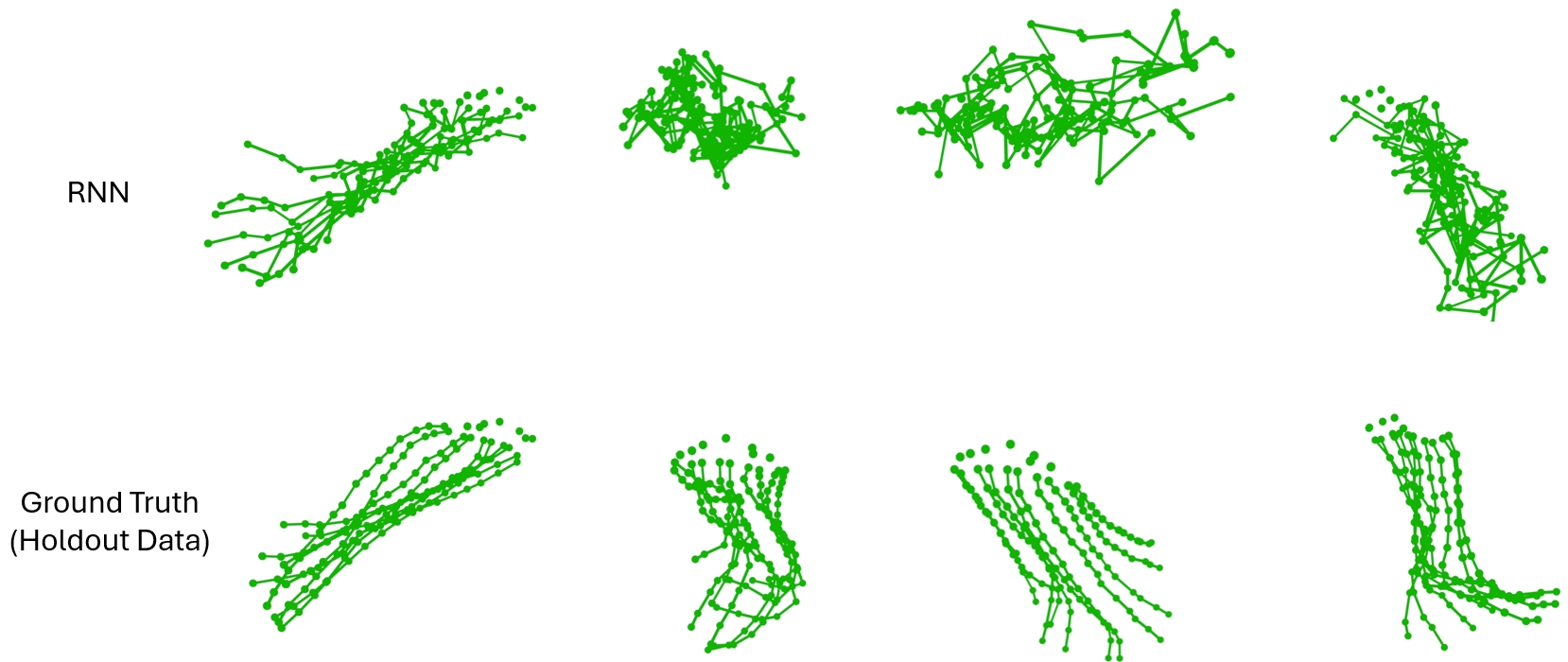}
    \caption{The main issue with the RNN is that it generalizes very poorly to the holdout data. Presumably, this issue could be fixed by collecting more and more training data, but that excessively increases the cost incurred in both data collection (via numerical simulation) and training (via optimization).  Even though our skinning and shape networks can smooth the noise in this RNN output (as they did in Figure \ref{fig:rnn_training}), the dynamics will still be completely wrong.}
    \label{fig:rnn_holdout}
\end{figure}

Finally, we consider an RNN approach. 
Following the low frequency module in \cite{pan2022predicting}, we trained a Gated Recurrent Unit (GRU) \cite{cho2014learning}. The inputs are the current body pose and the latent vector from the prior state, and the outputs are the current virtual bone positions and the current latent vector.
For the sake of a fair comparison, we used the same 5000 frames of (cape) simulation as training data; however, the RNN would obviously benefit from having access to an increased amount of training data.
A separate RNN was trained for each rope chain.
See Figures \ref{fig:rnn_training} and \ref{fig:rnn_holdout}.
Alternatively, the RNNs could be trained to match rope chain simulations; however, there is no reason to believe that predicting out-of-distribution data would fair any better than predicting in-distribution data.

\section{Conclusion and Future Work}
We presented a novel method for the real-time simulation of loose-fitting garments leveraging neural networks for both skinning and shape inference. We demonstrated that only a small number of degrees of freedom is necessary in order to capture the ballistic motions. In order to overcome the locking artifacts typically incurred by using only a small number of degrees of freedom, we proposed a rope chain simulation method that maintains inextensibilily while still allowing swinging, rotating, and buckling without resistance. A new history-based approach to collisions was introduced in order to accommodate fast-moving collision bodies and large time steps. A neural skinning solution was utilized in order to create a cloth mesh from the rope chain simulation degrees of freedom, and a quasistatic neural shape network was subsequently used in order to add additional details. These networks were trained with the aid of a PINN-style collision loss.

Since they are based on the PCA coefficients, the neural skinning and shape networks produce reasonable mesh output even with out-of-distribution input; however, out-of-distrubution input does adversely affect their ability to maintain an interpenetration-free state. 
Training the networks with a PINN-style collision loss gives reasonably penetration-free results for in-distribution data, but not necessarily for out-of-distribution data.
Improving the interpenetration- freeness for out-of-distribution input is perhaps the most important area for the future work.
Although the mesh could be post-processed to be interpenetration-free, doing so significantly decreases performance and thus the applicability for real-time applications. It is relatively cheap to process collisions for a small number of virtual bones, but far more expensive to process collisions for the entire mesh. Other interesting directions for future work include: making the rope chain simulation differentiable in order to automatically find constitutive parameters that best match the ground truth, investigating different network architectures for the neural skinning, generalizing a single network so that it can be used across different garment types and/or body shapes, etc.

\begin{acks}
Research supported in part by ONR N00014-19-1-2285, ONR N00014-21-1-2771, Epic Games, and Sony. We would like to thank Reza and Behzad at ONR for supporting our efforts into machine learning. We would also like to thank Michael Lentine for various insightful discussions.
\end{acks}

\bibliographystyle{ACM-Reference-Format}
\bibliography{reference}
\end{document}